\documentclass[1p,sort&compress]{elsarticle}
\usepackage{times}
\usepackage{latexsym,amsmath,amssymb,bm,euscript}
\usepackage[dvips]{color}
\usepackage{multirow}
\usepackage{bbm}
\usepackage{natbib}

\newcommand{\JJ}{\vec{\mathcal{J}}}
\newcommand{\QQ}{\vec{\mathcal{Q}}}
\newcommand{\FF}{\vec{\mathcal{F}}}
\newcommand{\GG}{\vec{\mathcal{G}}}
\newcommand{\curl}{\vec{\nabla}\times}
\newcommand{\divv}{\vec{\nabla}\cdot}
\def\ra{\rangle} % bra
\def\la{\langle} % ket

\def\br{{\bf r}}
\def\bR{{\bf R}}
\def\bx{{\bf x}}
\def\by{{\bf y}}
\def\bj{{\bf j}}
\def\bk{{\bf k}}
\def\cJ{{\cal J}}
\def\cQ{{\cal Q}}

\def\cG{{\cal G}}
\def\valpha{{\vec{\alpha}}}
\def\vbeta{{\vec{\beta}}}
\def\vgamma{{\vec{\gamma}}}
\def\half{{\frac{1}{2}}}
\def\Aext{A^{\rm ext}}
\def\vAext{{\vec{A}^{\rm ext}}}

\begin{document}

\title{Exact realization of Integer and Fractional Quantum Hall Phases in $U(1)\times U(1)$ models in $(2+1)d$}
\date{\today}
%\pacs{}

\author[scott]{Scott D. Geraedts}
\author{Olexei I. Motrunich}
\address{Department of Physics, California Institute of Technology, MC 149-33, Pasadena, California 91125, USA}
\cortext[scott]{Corresponding author. Email: sgeraedt@caltech.edu Phone: 626 319 9783}

%%%%%%%%%%%%%%%%%%%%%%%%%%%%%%%%%%%%%%%%%%%%%%%%%%%%%%%%%%%%%%%
\begin{abstract}
In this work we present a set of microscopic $U(1)\times U(1)$ models which realize insulating phases with a quantized Hall conductivity $\sigma_{xy}$.  The models are defined in terms of physical degrees of freedom, and can be realized by local Hamiltonians.  For one set of these models, we find that $\sigma_{xy}$ is quantized to be an even integer.  The origin of this effect is a condensation of objects made up of bosons of one species bound to a single vortex of the other species.  For other models, the Hall conductivity can be quantized as a rational number times two.  For these systems, the condensed objects contain bosons of one species bound to multiple vortices of the other species.  These systems have excitations carrying fractional charges and non-trivial mutual statistics.  We present sign-free reformulations of these models which can be studied in Monte Carlo, and we use such reformulations to numerically detect a gapless boundary between the quantum Hall and trivial insulator states.  We also present the broader phase diagrams of the models.
\end{abstract}

\maketitle

%%%%%%%%%%%%%%%%%%%%%%%%%%%%%%%%%%%%%%%%%%%%%%%%%%%%%%%%%%%%%%%%%%%%%%%%%

\section{Introduction}

A major project in condensed matter research is the discovery and classification of new topological phases.\cite{HasanKane_RMP, QiZhang_RMP, ChenGuLiuWen_Science, Chen2011, Chen2011b, KitaevTalk, LevinGu2012, LevinStern2012, LuVishwanath2012, Swingle2012, VishwanathSenthil2012, XuSenthil2013, Wen2013}  One class of these phases are the topological insulators.  Though much progress has been made in the study of topological insulators of free fermions,\cite{HasanKane_RMP, QiZhang_RMP} less is known about interacting systems.  
Much work has focused on understanding topological insulator phases that do not have long-range entanglement, also known as Symmetry Protected Topological (SPT) phases,\cite{ChenGuLiuWen_Science, Chen2011, Chen2011b, KitaevTalk} and very recent studies have also pursued cases with long-range entanglement.\cite{EssinHermele2012, MesarosRan2012, Maciejko2010}.  Recent works\cite{LuVishwanath2012, SenthilLevin2012, ChenWen2012, LiuWen2012} have used Chern-Simons approaches to provide understanding of interacting topological insulators in (2+1)D, and in particular of so-called integer Quantum Hall states of bosons (for a review, see Ref.~\cite{TurnerVishwanath2013}).  Several papers have proposed qualitative construction of such phases using Chern-Simons flux attachment\cite{SenthilLevin2012} and slave-particle approaches.\cite{GroverVishwanath2012, LuLee2012_QPT, LuLee2012_S1}  However, so far there have been no microscopic models producing such states.

Here we will present fully tractable realizations of both integer and fractional Quantum Hall phases with $U(1) \times U(1)$ symmetry.  Our approach is different from Chern-Simons and slave-particle approaches, in that we work directly with physical degrees of freedom and do not introduce artificial fluxes or enlarge the Hilbert space.  More specifically, we think directly in terms of charge and vortex degrees of freedom, which are all precisely mathematically defined in our lattice models.  In this way, our work is close in spirit to pursuits to understand fractionalized phases of spins and bosons in terms of the vortex physics.\cite{BalentsFisherNayak1999, SenthilFisher_Z2}  The $U(1) \times U(1)$ structure allows us to provide an unambiguous and simple physical picture of the integer and fractional quantum Hall states of bosons.  Thus, an elementary integer quantum Hall state is obtained when a vortex in one species binds a charge of the other species and the resulting composite object condenses; our approach provides a precise meaning of such a condensation.  General integer quantum Hall states are obtained when a vortex in one species binds a fixed number of charges of the other species.  On the other hand, fractional quantum Hall states are obtained when we have condensation of composite objects that are bound states of $d$ vortices of one species and $c$ particles of the other species.  In this case we show that the system has a fractional quantum Hall response given by $\sigma_{xy} = 2c/d$ and has quasiparticles carrying fractional charges of $1/d$ of the microscopic charges and non-trivial mutual statistics.  Such $d$-tupled vortex condensation leading to charge fractionalization is reminiscent of the idea in Refs.~\cite{BalentsFisherNayak1999, SenthilFisher_Z2} of paired-vortex condensation leading to $Z_2$ fractionalized phases, although in the present case there is also binding of charges and the phase shows quantized Hall response.
The integer quantum Hall phases discussed in this work are examples of SPT phases, while the fractional quantum Hall states are examples of interacting topological insulators with topological order and long-range entanglement.

An interesting aspect of our models is that they are examples of interacting topological insulators that can be studied in a sign-free Monte Carlo, thus adding to a growing list of exotic quantum phases and transitions that can be explored numerically.\cite{KaulMelkoSandvik2012}  Furthermore, perhaps for the first time, we can simulate systems that have non-trivial quantum Hall responses and in particular gapless edge states.  In this paper, we perform such a study of a boundary between Quantum Hall and trivial insulators and provide compelling evidence of gaplessness at the boundary.

The paper is organized as follows.  In Sec.~\ref{sec::demon} we describe our models and demonstrate that they realize integer and fractional quantum Hall phases.  In Sec.~\ref{sec:reform} we derive the sign-free reformulations used in the numerics.  In Sec.~\ref{sec::results} we present the results of these numerics, including evidence for gapless modes at the boundary of the quantum Hall states.  In Sec.~\ref{sec:reverse} we show the broader phase diagrams for our models.  \ref{app:duality} reviews a duality transformation which is used throughout this paper. In \ref{app:connections} we demonstrate the connections between our models and various field-theoretic approaches, specifically non-linear sigma models with topological terms and $K$-matrix theories.  Finally, in \ref{app:H} we present a local Hamiltonian which gives the action discussed in the main text.

%%%%%%%%%%%%%%%%%%%%%%%%%%%%%%%%%%%%%%%%%%%%%%%%%%%%%%%%%%%%%%%%%%%%%%%%
\section{Explicit Models with Integer and Fractional Quantum Hall Effect}
\label{sec::demon} 

In this section we introduce the models which are the subject of this work. We then rewrite the models in terms of new variables, in the process demonstrating that the models realize quantum Hall phases.
The models are given by the following action in (2+1)D Euclidean space-time:
\begin{eqnarray}
S&=&\frac{1}{2} \sum_{r,r'} v_1(r-r')\JJ_1(r)\cdot\JJ_1(r')
+\frac{1}{2}\sum_{R,R'} v_2(R-R')\JJ_2(R)\cdot\JJ_2(R')\nonumber\\
&+&i\sum_{R,R'}  w(R-R')[\vec{\nabla}\times\JJ_1](R)\cdot\JJ_2(R').
\label{action}
\end{eqnarray}
The $\mathcal{J}_1$ variables are integer-valued conserved currents residing on a cubic lattice (the `direct' lattice) labeled with the index $r$. They satisfy $\vec{\nabla}\cdot\JJ_1=0$, which means that they form closed loops. The $\mathcal{J}_2$ variables are the same except that they reside on a lattice dual to the direct lattice, and whose sites are labeled by the index $R$. 
These variables represent space-time currents of two separately conserved species of bosons.
In (2+1)D, applying the curl operator to a vector object on the direct lattice, such as a $\JJ_1(r)$ variable, leads to a vector object on the dual lattice, denoted $[\curl\JJ_1](R)$.

The first two terms in the above action are short-ranged repulsive interactions within each species of boson. The third term is a short-ranged interaction between currents of different species. The latter term may appear unusual, but this action is local, and in \ref{app:H} we show that it can arise from a local Hamiltonian.  Also, even though it appears that this term treats the $\cJ_1$ and $\cJ_2$ currents differently, we can integrate by parts and see that, as long as $w(R,R') = w(R-R') = w(R'-R)$, the $\cJ_1$ and $\cJ_2$ enter symmetrically.

In loop actions such as Eq.~(\ref{action}), it is possible to make a change of variables to obtain equivalent actions that may be easier to interpret.  One such change of variables is a well-known duality transformation, (given in \ref{app:duality}) which can be generalized\cite{Gen2Loops} to arbitrary modular group transformations.\cite{Cardy1982, Shapere1989}
This is the change of variables we will use to show that our model can produce integer and fractional bosonic quantum Hall insulators.

To make this change of variables, we first rewrite Eq.~(\ref{action}) in $k$-space as follows:
\begin{eqnarray}
S&=&\frac{1}{2}\sum_k \left[v_1(k)|\JJ_1(k)|^2+v_2(k)|\JJ_2(k)|^2\right]
+i\sum_k \theta(k)\JJ_1(-k)\cdot \vec{a}_{\mathcal{J}2}(k).
\label{kaction}
\end{eqnarray}
In the above expression, $\JJ_1(k) = \sum_r \JJ_1(r)e^{-i\vec{k}\cdot\vec{r}} / \sqrt{\rm Vol}$ and $v_1(k)=\sum_r v_1(r-r')e^{-i\vec{k}\cdot(\vec{r}-\vec{r'})}$.  $\JJ_2(k)$ and $v_2(k)$ are defined similarly to $\JJ_1(k)$ and $v_1(k)$, but on the dual lattice indexed by $R$. The ``gauge'' field $\vec{a}_{\mathcal{J}2}(r)$ is defined such that $\JJ_2(R)=[\vec{\nabla}\times\vec{a}_{\mathcal{J}2}](R)$ (such representation is possible given the divergenceless conditions and also conditions of the vanishing total currents discussed below, and the action is independent of the gauge choice\cite{Loopy, short_range3}).  Starting from Eq.~(\ref{action}), we obtain $\theta(k)=|\vec{f}_k|^2\sum_R  w(R-R')e^{-i\vec{k}\cdot(\vec{R}-\vec{R'})}$, where $|\vec{f}_k|^2\equiv\sum_\mu (2-2\cos{k_\mu})$ arises from the lattice derivatives and the sum is over all directions. We assume that $v_1(k)$, $v_2(k)$ and $\theta(k)$ are even in $k$.

In order to determine the physical properties of the model, we couple it to external gauge fields by adding the following terms to Eq.~(\ref{kaction}):
\begin{equation}
\delta S=i\sum_k \left[ \JJ_1(-k)\cdot \vec{A}_{1}^{\rm ext}(k) + \JJ_2(-k)\cdot \vec{A}_{2}^{\rm ext}(k) \right].
\label{Aext}
\end{equation}

We can use the duality transform from \ref{app:duality} to go from the $\cJ_1$ variables to dual $\cQ_1$ variables as follows:
\begin{eqnarray}
&S&=\frac{1}{2}\sum_k \frac{\left|2\pi\QQ_1(k)+\theta(k)\JJ_2(k)+[\curl\vec{A}_{1}^{\rm ext}](k)\right|^2}{|\vec{f}_k|^2v_1(k)}\nonumber\\
&+&\frac{1}{2}\sum_k v_2(k)|\JJ_2(k)|^2 + i \sum_k \JJ_2(-k) \cdot \vec{A}_{2}^{\rm ext}(k).
\label{JQ1}
\end{eqnarray}
If we interpret the $\mathcal{J}_1$ variables as physical bosons, the $\mathcal{Q}_1$ variables can be interpreted as vortices in the boson phase variables. Like the $\mathcal{J}_1$ variables, the $\mathcal{Q}_1$ variables are divergenceless and therefore form closed loops.
A technical remark: We have found it convenient to require zero total current in our system, $\JJ_{1{\rm tot}} \equiv \sum_r \JJ_1(r) = 0$ and $\JJ_{2{\rm tot}}=0$. This makes the above duality procedure exact,\cite{Loopy, short_range3} and requires $\QQ_{1{\rm tot}}=0$. All of the currents defined below will also satisfy this condition.

As discussed in our previous work,\cite{Gen2Loops} in the above action we can make the following change of variables:
\begin{eqnarray}
\FF_1 &=& a\QQ_1 - b\JJ_2,\\
\GG_2 &=& c\QQ_1 - d\JJ_2.
\label{modularshift}
\end{eqnarray}
This change of variables is valid if the matrix
\begin{equation}
\begin{pmatrix}
a & b \\
c & d 
\end{pmatrix}
\in PSL(2,\mathbb{Z}),
\end{equation}
i.e.,  $a,b,c,d$ are integers such that $ad-bc=1$. Since the above matrix is an element of the modular group, we call this change of variables a modular transformation and will often refer to it simply $(a,b,c,d)$. We can then perform the duality transform to go from the $\mathcal{F}_1$ variables to dual $\cG_1$ variables, which gives us an action in terms of the $\mathcal{G}_1$ and $\cG_2$ variables. This transformation, from $\cJ_1$, $\cJ_2$ variables to $\cG_1$, $\cG_2$ variables, is the generalization of the duality operation to modular transformations. After performing this change we are left with the following action:
\begin{eqnarray}
S&=&\frac{1}{2}\sum_k v_{\mathcal{G}1}(k) \left|\GG_1(k)+\frac{c[\curl\vec{A}_{2}^{\rm ext}](k)}{2\pi}\right|^2
+\frac{1}{2}\sum_k v_{\mathcal{G}2}(k)\left|\GG_2(k)+\frac{c[\curl\vec{A}_{1}^{\rm ext}](k)}{2\pi}\right|^2\nonumber\\
&+&i\sum_k \theta_\mathcal{G}(k) \GG_1(-k)\cdot \vec{a}_{\cG2}(k)
-i\sum_k \frac{c[2\pi a-\theta_\mathcal{G}(k)c]}{(2\pi)^2} [\curl\vec{A}_{1}^{\rm ext}](-k)\cdot \vec{A}_{2}^{\rm ext}(k)\nonumber\\
&-& i\sum_k \left[a - \frac{\theta_\mathcal{G}(k) c}{2\pi}\right]\left[\GG_1(-k) \cdot \vec{A}_{1}^{\rm ext}(k) + \GG_2(-k) \cdot \vec{A}_{2}^{\rm ext}(k) \right],
\label{gaction}
\end{eqnarray}
where $\GG_2=\vec{\nabla}\times \vec{a}_{\cG2}$ and
\begin{eqnarray}
&&\!\!\!\!\! v_{\mathcal{G}1/2}(k)=\frac{(2\pi)^2v_{1/2}(k)}{[2\pi d+\theta(k) c]^2+ v_{1}(k)v_{2}(k)|\vec{f}_k|^2c^2},\label{VJ}\\
&&\!\!\!\!\! \frac{\theta_\mathcal{G}(k)}{2\pi}=\frac{[2\pi b+\theta(k) a][2\pi d + \theta(k) c]+v_{1}(k)v_{2}(k)|\vec{f}_k|^2ca}{[2\pi d+\theta(k) c]^2+ v_{1}(k)v_{2}(k)|\vec{f}_k|^2c^2}.\nonumber
\end{eqnarray}

We can deduce the quantum Hall properties of the system from the action in terms of the $\cG$ variables. If we ignore the external fields $\vec{A}^{\rm ext}_{1/2}$, we see that this action has the same form as Eq.~(\ref{kaction}), but the action is defined in terms of different variables which see different inter-particle potentials.  If $v_\cG$ is large, so that only small loops in the $\mathcal{G}$ variables can form, we say that the $\mathcal{G}$ variables are `gapped', and in this phase we can interpret the $\mathcal{G}$ variables as describing gapped quasiparticles above the ground state.  Equation (\ref{gaction}) is the action for these quasiparticles.  Restoring the external fields  $\vec{A}^{\rm ext}_{1/2}$ and integrating out the gapped $\mathcal{G}$ variables, we would only generate Maxwell-like terms for the external fields (assuming that $v_\cG$ potentials are short-ranged, which is true for $d\neq 0$).  Hence, we can extract Hall-like conductivity describing transverse-cross response from the fourth term in the above action:
\begin{equation}
\sigma_{xy}^{12}(k)=\frac{2c[2\pi a-\theta_\cG(k)c]}{2\pi}.
\end{equation}
This conductivity is defined in units of $e^2/h$. 

Now consider the conductivity in the $k\rightarrow0$ limit. Here and below we assume $d \neq 0$.  (If we use the above transformation with $d=0$, the $\cG$ particles are essentially vortices in the $\cJ$, and if they are gapped, this gives a superfluid and will be discussed in Sec.~\ref{sec:reverse}.)   From Eq.~(\ref{VJ}) we can see that in this limit and for short-ranged $v(k)$, 
\begin{equation}
\theta_{\mathcal{G}}=2\pi \frac{b}{d}+O(k^2).
\end{equation}
This describes the mutual statistics of the $\cG_1$ and $\cG_2$ particles.
We can use  $\theta_{\mathcal{G}}$ to determine the conductivity in the $k\rightarrow0$ limit: 
\begin{equation}
\sigma_{xy}^{12}=2\frac{c}{d}.
\label{sigma}
\end{equation}
We see that for short-ranged potentials, we have a universal, rational conductivity.  From the last term in Eq.~(\ref{gaction}) we can extract the charges of the $\cG_1$ and $\cG_2$ quasiparticles to be
\begin{equation}
{\rm charge} = \frac{1}{d}, \label{charge1d}
\end{equation}
relative to $\Aext_1$ and $\Aext_2$ respectively.
We see that when $d=1$, the system has a Hall conductivity quantized to be an even integer and excitations carrying integer charges. We propose that this is a realization of the bosonic integer quantum Hall effect.\cite{LuVishwanath2012}  When $d>1$, we see that the Hall conductivity is quantized as a rational number and the $\mathcal{G}$ quasiparticles carry fractional charge. Therefore this phase is a fractional quantum Hall effect for bosons. 

The above results are true only in the region where the $\cal{G}$ variables are gapped. For a general set of potentials $v_{1/2}(k),\theta(k)$ and coefficients $a,b,c,d$, we must use numerics to determine whether such a gapped region exists, and where it can be found. We now present a simple, short-ranged model in which a gapped region always exists and we know where the gapped region is located. This model also has the feature that the above statistical angle $\theta_\cG(k)$ and charges of the quasiparticles are independent of $k$.  This will simplify our analysis, though we expect the results we derive for these specific choices will hold also for any other short-ranged potentials in the $k\rightarrow0$ limit, as long as the $\cG$ particles are gapped.  (In \ref{app:connections}, we show how the above analysis can be viewed as a derivation of an effective $K$-matrix-like theory at long wavelengths for our quantum Hall phases.)

To get these simple models we choose:
\begin{eqnarray}
&&v_{1/2}(k)=\frac{\lambda_{2/1}}{\lambda_1\lambda_2+\frac{c^2|\vec{f}_k|^2}{d^2(2\pi)^2}}, \label{vintro}\\
&&\theta(k)=\frac{-c}{2\pi d}\frac{|\vec{f}_k|^2}{\lambda_1\lambda_2+\frac{c^2|\vec{f}_k|^2}{d^2(2\pi)^2}},\label{tintro}
\end{eqnarray}
where $\lambda_1$ and $\lambda_2$ are real-valued parameters. The potentials also contain parameters $c$ and $d$ which are integer-valued. If we choose the coefficients $c$ and $d$ in the transformation Eq.~(\ref{modularshift}) [leading to Eq.~(\ref{VJ})] to be the same as those in these potentials, the resulting action for the $\cal{G}$ variables will have the simplifying features described above. As we will see in a moment, in this case the parameters $c$ and $d$  give the conductivities and fractionalized charges of the quantum Hall phases occurring at small $\lambda_{1,2}$, and therefore they are useful labels of the resulting quantum Hall phases. We will explain how these potentials were determined in Sec.~\ref{sec:reverse}. 

With the above choices, the potentials in the action in terms of the $\mathcal{G}$ variables take the following simple form:
\begin{eqnarray}
v_{\mathcal{G}i}(k)&=&\frac{1}{d^2\lambda_i},\label{vgd2} \\
\theta_{\mathcal{G}}&=&2\pi\frac{b}{d}.\label{tg}
\end{eqnarray}
Therefore we can easily see that with the above choices we will always have a phase where the $\mathcal{G}_i$ variables are gapped in the limit of small $\lambda_i$. In this limit the system will be a bosonic quantum Hall insulator.

If we insert the above expressions for $v(k)$ and $\theta(k)$ into Eq.~(\ref{JQ1}), we see that we can write that action in the following way:
\begin{eqnarray}
S &=& \frac{1}{2}\sum_{k} \frac{(2\pi)^2\lambda_1}{|\vec{f}_k|^2 } |\QQ_1(k)|^2\label{JQreal}
+ \frac{1}{2}\sum_{R} \frac{1}{\lambda_2} |\JJ_2(R) - \eta(R)\QQ_1(R)|^2.
\end{eqnarray}
Here $\eta(R)=c/d$ everywhere in the system, though later we will consider spatially varying $\eta(R)$. Studying this action at small $\lambda_1$ and $\lambda_2$ can tell us about the physics of the quantum Hall states.

\begin{figure}
\begin{center}
\includegraphics[width=0.5\linewidth]{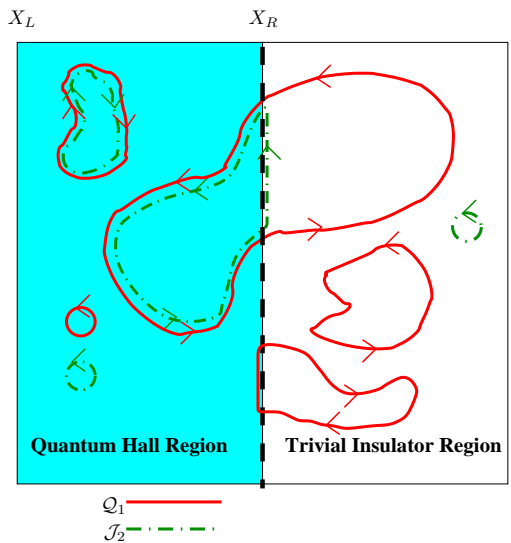}%
\caption{A sketch of the situation where $\eta(R)$ is allowed to vary in space, with $\eta(R)=1$ in the quantum Hall region and $\eta(R)=0$ in the trivial insulator region. In the trivial insulator region, loops of the $\mathcal{Q}_1$ variables can be large, while the $\mathcal{J}_2$ variables can only form small loops. In the quantum Hall region, large loops of composite objects with ($\mathcal{Q}_1=1$, $\mathcal{J}_2=1$) exist, while loops of only $\mathcal{Q}_1$ or only $\mathcal{J}_2$ variables are small. In order to have closed loops of the $\mathcal{J}_2$ variables, there must be behavior on the edge different from that in the bulk, which leads to the gapless edge supporting non-zero Hall conductivity.}
\label{loops}
\end{center}
\end{figure}

First consider the situation where $c=0$, which leads to $\eta(R)=0$, and consider the limit of small $\lambda_1$, $\lambda_2$. We can see from Eq.~(\ref{sigma}) that $\sigma^{12}_{xy}$ vanishes and therefore this system is not in a Quantum Hall state. 
There is a small energy cost for loops in the $\mathcal{Q}_1$ variables, but a large energy cost for loops in the $\mathcal{J}_2$ variables. Therefore the $\mathcal{J}_2$ loops are gapped. The $\mathcal{Q}_1$ variables are `condensed', which means that large loops of these variables can form. In general, since $\mathcal{J}_1$ and $\mathcal{Q}_1$ are related by a duality transformation, if one of them is gapped the other is condensed. Therefore the $\mathcal{J}_1$ variables are gapped in this phase. Since the physical $\mathcal{J}_1$ and $\mathcal{J}_2$ variables are gapped, the system is a trivial insulator. This situation is shown on the right side of Fig.~\ref{loops}.  We could have arrived at this conclusion also more quickly by examining Eqs.~(\ref{vintro})-(\ref{tintro}) for $c=0$, but it was convenient to develop the picture in the $\cQ_1$ and $\cJ_2$ variables.

Now consider the case where $c\neq0$, so that $\eta(R)=c/d$. Since the $\mathcal{G}$ variables are gapped we have a quantized Hall conductivity, and the system is in the bosonic quantum Hall state.
Notice that composite objects with $\mathcal{J}_2=c$ and $\mathcal{Q}_1=d$ see a very small energy cost, and therefore large loops of such objects can form.  On the other hand, both the $\mathcal{J}_2$ and $\mathcal{Q}_1$ variables see large potentials if they exist independently, so only small loops of these variables can form by themselves. This is illustrated on the left side of Fig.~\ref{loops}.

We would like to know what carries the charges that leads to the $\sigma^{12}_{xy}\neq 0$. By analogy with the fermionic Quantum Hall effect, we expect that the charges are being carried by edge states. We will examine the physics of the formation of edge states by including a boundary between the quantum Hall state and the trivial insulator in our system.
This is accomplished by allowing $\eta(R)$ to vary in space. Therefore we have one region where $\eta(R)=c/d$, and we have another region where $\eta(R)=0$.  Note that we have defined $\eta(R)$ as varying on the dual lattice denoted by the index $R$. Before we allowed $\eta$ to vary in space, we had a symmetry between the $\mathcal{J}_1$ and $\mathcal{J}_2$ variables in the case where $v_1=v_2$. However the different variables see the boundary differently and it  breaks this symmetry.

Now consider what happens at the boundary of the quantum Hall state. For example, consider the case where $\eta(R)=1$ in the region where it is non-zero. In the quantum Hall region we will have large loops with ($\mathcal{J}_2=1$,$\mathcal{Q}_1=1$), while in the trivial insulating region we have large loops of only $\mathcal{Q}_1$ variables. This situation is illustrated in Fig.~\ref{loops}. The large loops of the $\mathcal{Q}_1$ variables can pass through the boundary. However, the $\mathcal{J}_2$ loops must be bound to the $\mathcal{Q}_1$ loops in one region and must disappear in the other region. The system must find a way to accomplish this while satisfying the constraint that all currents must be divergenceless. 
To do this, it exhibits behavior which would be energetically forbidden in the bulk. For instance, the $\cJ_2$ currents could run along the edge, as seen in Fig.~\ref{loops}. In this case on the right edge, the $\cJ_2$ currents run from places where the $\cQ_1$ variables cross the boundary from left to right to places where they cross from right to left. Alternatively, loops of the $\cQ$ variables could be forbidden from crossing between the two regions, which means that vortices in the boson phase variables would be forbidden on the boundary. This unusual behavior leads to gapless modes on the boundary between a quantum Hall region and a trivial insulator. In Sec.~\ref{sec::results} we will develop mathematical description of this behavior and will numerically find evidence for these gapless modes.

%%%%%%%%%%%%%%%%%%%%%%xfig output for loops figure

%%%%%%%%%%%%%%%%%%%%%%%%%%%%%%%%%%%%%%%%%%%%%%%%%%%%%%%%%%%%%%%%%%%%%%%%
\section{Monte Carlo Study of a Model with a Boundary}
\label{sec:reform}
In order to further characterize these bosonic quantum Hall phases, we now seek to find gapless edge modes numerically using Monte Carlo simulations.\cite{Senthil_Ashvin_thanks}  We include edges by allowing $\eta(R)$ to vary in space, in one direction, which we will label as the $x$-direction. In our large $L\times L\times L$ system, we will say that 
\begin{equation}
\eta(R)=\left\{\begin{array}{cc}
c/d & {\rm for}~ X_L \leq X < X_R \\
0 & {\rm otherwise}
\end{array}\right. ,
\label{etaspace}
\end{equation}
where $X$ is the $x$-coordinate on the dual lattice. The quantum Hall behavior occurs in the region with non-zero $\eta$. 
We will take half of the system to be in the quantum Hall phase and the other half in the trivial insulator phase.

Note that the original action in terms of $\cJ_1, \cJ_2$ currents, Eq.~(\ref{action}), has a sign problem.  On the other hand, the action in terms of $\cQ_1, \cJ_2$, Eq.~(\ref{JQreal}), is a statistical mechanics model that can be directly studied in Monte Carlo. However, we can see that the $\mathcal{Q}_1$ variables have long-ranged interactions, and in systems with such interactions the time required for the simulations scales as $L^6$. We would like to have only short-ranged interactions in our system, so that the Monte Carlo can run in a time proportional to $L^3$, and larger sizes can be studied. The remainder of this section will be devoted to developing several exact reformulations of Eq.~(\ref{JQreal}) which have only short-ranged interactions. These reformulations are also useful because they provide access to boson Green's functions.

Consider the following action:
\begin{eqnarray}
&&S[\vec{\alpha}_1,\vec{\alpha}_2,\JJ_1,\JJ_2]=\frac{1}{2}\sum_r \lambda_1 \vec{\alpha}_{1}(r)^2 + \frac{1}{2}\sum_R \lambda_2 \vec{\alpha}_{2}(R)^2  \\
&&~~~ +i\sum_R \frac{\eta(R)}{2\pi} [\curl\vec{\alpha}_{1}](R) \cdot \vec{\alpha}_{2}(R)
 +i\sum_r \JJ_1(r)\cdot\vec{\alpha}_{1}(r)+i\sum_R \JJ_2(R)\cdot\vec{\alpha}_{2}(R).\nonumber
\label{preal}
\end{eqnarray}
The new $\alpha_1$ and $\alpha_2$ variables are real-valued vector fields residing on links of the direct and dual lattices respectively. One can check that the partition sum which results from this action is the same as Eq.~(\ref{action}) if we integrate out both of these species of vector fields.  Note that the $\alpha_1$ and $\alpha_2$ variables are not some gauge fields; rather, they are some local massive fields and are integrated over with no constraints.  \ref{app:H} shows that such an action can be realized as a path integral for a local Hamiltonian with bosons coupled to oscillator degrees of freedom.  \ref{app:connections} motivates the connection with Eq.~(\ref{action}) by starting from an action in terms of vortex variables, while we can also derive the result Eq.~(\ref{JQreal}) for general $\eta(R)$ from a direct analysis below.

In order to get a sign-free action that we can study efficiently in Monte Carlo, we start with Eq.~(\ref{preal}) but perform different integrations. 
First, we integrate out the $\alpha_{2}$ variables. Then, we would like to integrate out the $\mathcal{J}_1$ variables. However, the $\mathcal{J}_1$ variables are integer-valued and constrained to be divergenceless with no total current. We enforce these constraints by adding new variables to our partition sum as follows.
To enforce the divergenceless of the $\mathcal{J}_1$ variables we add the following term:
\begin{equation}
\delta_{[\divv\JJ_1](r)=0}=\int_{-\pi}^{\pi} d\phi_1(r) \exp\left[-i\phi_1(r)[\divv\JJ_1](r)\right].
\label{zphi}
\end{equation}
(We are ignoring overall constants here and below.) This introduces a $2\pi$-periodic $\phi_1(r)$ variable on every site of the lattice. These variables correspond to the phases of the type-1 bosons. We enforce the constraint that there must be no total current (in our full system) by adding another term to the partition sum:
\begin{eqnarray}
\delta_{\JJ_{1{\rm tot}}=0}=\prod_{\mu=1}^{3}\int_{-\pi}^{\pi} d\gamma_{1\mu} \exp[-i\gamma_{1\mu}\sum_r \delta_{r_\mu=0}\mathcal{J}_{1\mu}(r)].
\label{zgamma}
\end{eqnarray}
This term introduces a $2\pi$-periodic $\gamma_{1\mu}$ variable for each direction $\mu$ on the lattice. This variable means that instead of periodic boundary conditions we have a fluctuating boundary condition such that across the boundary the $\phi$ variables differ by $\gamma$; here we chose the boundary plane perpendicular to the $x$ direction to be at $x=0$, and similarly for the other directions. Note that these boundary terms are not related to the edges between quantum Hall and trivial insulator regions in the system.  They are only included to make the duality procedure between the $\cJ$ and $\cQ$ variables precise, so that we are confident that our various reformulations simulate exactly the same system for any finite size.

Now that the $\mathcal{J}_1$ variables are unconstrained, we can go from integer-valued $\mathcal{J}_1$ to real-valued $j_1$ by using the following relation:
\begin{equation}
\sum_{\mathcal{J}_{1\mu}(r)}[...]=\int_{-\infty}^{+\infty}\!\! dj_{1\mu}(r) \sum_{p_{1\mu}(r)} \exp\left[-i2\pi p_{1\mu}(r)j_{1\mu}(r)\right][...].
\label{zp}
\end{equation} 
This term introduces integer-valued $p_{1\mu}(r)$ variables on every link on the lattice, and these variables are free of any constraint.  In formal duality maps,\cite{PolyakovBook, Peskin1978, Dasgupta1981, FisherLee1989, LeeFisher1989, artphoton, short_range3} the physical meaning of the $p_1$ variables is that their curl gives vorticity in the phase variables conjugate to $\mathcal{J}_1$, i.e., $\QQ_1 = \vec{\nabla} \times \vec{p}_1$.

With all the constraints implemented, we can now integrate out the $j_1$ variables to get:
\begin{eqnarray}
&&Z=\sideset{}{'}\sum_{\JJ_2}  \sum_{\vec{p}_1}  \int_{-\pi}^{\pi}  \mathcal{D}\phi_1 \int_{-\pi}^{\pi} \prod_{\mu=1}^{3} d\gamma_{1\mu}
e^{-S[\phi_1,\gamma_1,\vec{p}_1,\JJ_2]} , ~~~~~
\label{Zint} \\
&&~~ S[\phi_1,\gamma_1,\vec{p}_1,\JJ_2] =
\frac{\lambda_1}{2}\sum_{r} \left[\vec{\omega}_{1}(r)-2\pi \vec{p}_1(r) \right]^2 \nonumber\\
&&~~~~~~~~ + \frac{1}{2\lambda_2}\sum_R \left[\JJ_2(R)-\eta(R)(\curl\vec{p}_1)(R) \right]^2.
\label{Sfrac1}
\end{eqnarray}
Here the primed sum over $\cJ_2$ variables is subject to the constraints mentioned earlier; we have also defined
\begin{equation}
\omega_{1\mu}(r)=\phi_1(r+\hat{\mu})-\phi_1(r)-\gamma_{1\mu}\delta_{r_\mu=0},
\label{omega}
\end{equation}
and used $\curl\vec{\omega}_1 = 0$.
This action is sign-free and all interactions are short-ranged, so it can be studied efficiently in Monte Carlo; it will allow us to detect gaplessness of the quantum Hall edge by looking at the spatial correlations of the $\phi_1$ variables. The action is not explicitly $2\pi$-periodic in the $\omega_1$ variables, but all physical measurements have the required periodicity.  Specifically, the periodicity $\phi_1(r) \to \phi_1(r) + 2\pi N(r)$ and $\gamma_{1\mu} \to \gamma_{1\mu} + 2\pi M_\mu$, with integers $N(r)$ and $M_\mu$, can be accounted for by shifting the summation variables $p_{1\mu}(r) \to p_{1\mu}(r) + \nabla_\mu N(r) - M_\mu \delta_{r_\mu = 0}$, which does not change $\curl\vec{p}_1$ in the second term of Eq.~(\ref{Sfrac1}). 
We will present Monte Carlo measurements of the $e^{i\phi_1}$ correlators at the edge in the next section.
We can also use Eq.~(\ref{Sfrac1}) to directly show equivalence with the action Eq.~(\ref{JQreal}).  Indeed, if we separate the $\phi_1, \vec{p}_1$ system into spin-wave part and vortex part $\QQ_1 = \curl\vec{p}_1$, and integrate out the spin-wave part (see e.g., \ref{app:duality}), we would obtain exactly Eq.~(\ref{JQreal}).

We would also like to measure Greens functions of the type-2 bosons.  To get a reformulation which provides access to the phase variables of the type-2 bosons, $\phi_2$, we go back to Eq.~(\ref{preal}) and integrate by parts on the second line as follows:
\begin{eqnarray}
 \sum_R \eta(R) [\curl\vec{\alpha}_{1}](R) \cdot \vec{\alpha}_{2}(R) 
= \sum_r (\curl[\eta(R)\vec{\alpha}_2(R)])(r)\cdot \vec{\alpha}_1(r) ~,
\end{eqnarray}
which already suggests that the two boson species will see the edge differently.
We then integrate out the $\alpha_1$ and $\mathcal{J}_2$ variables using similar methods to those above. The resulting action is:
\begin{eqnarray}
&&S[\phi_2,\gamma_2,\vec{p}_2,\JJ_1]=
\frac{\lambda_2}{2}\sum_R [\vec{\omega}_{2}(R)-2\pi \vec{p}_{2}(R)]^2\nonumber\\
&& ~~~ +\frac{1}{2\lambda_1}\sum_r \bigg[\JJ_1(r)-(\curl[\eta(R)\vec{p}_2(R)])(r)
+ \frac{1}{2\pi}(\curl[\eta(R)\vec{\omega}_2(R)])(r) \bigg]^2 , ~~~~
\label{Sfrac2}
\end{eqnarray}
where $\omega_{2\mu}(R)$ is defined similarly to Eq.~(\ref{omega}). For the $\eta(R)$ in Eq.~(\ref{etaspace}) we can calculate:
\begin{eqnarray}
(\curl[\eta(R)\vec{\omega}_{2}(R)])(r) 
 = \frac{c}{d}(\delta_{x+\frac{1}{2}=X_L}-\delta_{x+\frac{1}{2}=X_R})\times
[\omega_{2y}(R)\hat{z}
-\omega_{2z}(R)\hat{y}],
\end{eqnarray}
which is non-zero only near the boundaries.  In the last line $\omega_{2z}(R)$ is at the appropriate edge ($X=X_R$ or $X=X_L$) and is calculated on the dual lattice link perpendicular to the direct lattice link $\langle r, r+\hat{y}\rangle$ next to it, and similarly for $\omega_{2y}(R)$.
Thus we have the extra boundary $\omega_2$ term in Eq.~(\ref{Sfrac2}), which comes from differentiating $\eta(R)$. 

Note that the $2\pi$-periodicity in the $\omega_2$ variables is accounted for by the $p_2$ variables.  A more subtle observation is that in the limit of $\lambda_2 = 0$, the $\phi_2$ variables have periodicity of $2\pi/c$.  Indeed, if we shift $\phi_2(R)$ by $2\pi N(R)/c$ with integer-valued $N(R)$, then $\vec{\omega}_2(R)$ is shifted by $2\pi \vec{\nabla}N(R)/c$.  We can simultaneously shift the summation variables $\vec{p}_2(R)$ by $-b\vec{\nabla}N(R)$ and observe that $[\vec{\omega}_2(R) - 2\pi \vec{p}_2(R)] \eta(R)/(2\pi)$ is shifted by $\vec{\nabla}N(R) (ad/c)\eta(R)$, where we used $ad - bc = 1$.  The latter shift is an integer-valued vector field [since $(ad/c)\eta(R)$ is an integer everywhere], and its curl can be absorbed into the redefinition of $\JJ_1(r)$, thus keeping the $\lambda_1$ term in Eq.~(\ref{Sfrac2}) unchanged.  We will need to keep this in mind when measuring correlation functions of the $\phi_2$ variables.

In the case where $d=1$ so that $\eta(R)$ is an integer everywhere, we can obtain one additional reformulation from Eq.~(\ref{Sfrac2}). Consider making the following change of variables:
\begin{equation}
\vec{M}(r)=\JJ_1(r)-(\curl[\eta(R)\vec{p}_2(R)])(r).
\end{equation}
This is an allowed change of variables since $\eta(R)\vec{p}_2(R)$ is an integer vector field and therefore its curl is a divergenceless integer field.
Note that if $\eta(R)$ is not an integer, such a change of variables is not allowed since it will lead to a non-integer $\vec{M}$. After making this change of variables, we can perform a summation over $\vec{p}_2$ to arrive at the following action:
\begin{eqnarray}
&& \hspace{-25pt} S[\phi_2,\gamma_2,\vec{M}]=
\sum_{R,\mu} V_{\rm Villain}[\omega_{2\mu}(R);\lambda_2] 
\label{Sint}
+\frac{1}{2\lambda_1}\sum_r\left[\vec{M}(r)+\frac{1}{2\pi}(\curl[\eta(R)\vec{\omega}_{2}(R)])(r)\right]^2 \\
&& \exp[-V_{\rm Villain}(\theta;\lambda)]=\sum_{p=-\infty}^{+\infty} \exp\left[-\frac{\lambda}{2}(\theta-2\pi p)^2\right] .
\end{eqnarray}
We tabulate the Villain potential in the last line before the start of the simulations.  Since this action contains fewer variables than the previous actions, it is more efficient to run in Monte Carlo.

%%%%%%%%%%%%%%%%%%%%%%%%%%%%%%%%%%%%%%%%%%%%%%%%%%%%%%%%%%%%%%%%%%%%%%%%
\section{Numerical Evidence for Gapless Edge}
\label{sec::results}
In order to determine the existence of gapless modes on the edge, we will measure correlators of the $\phi_1$ and $\phi_2$ variables. 

A technical point: 
these variables are not translationally invariant because of the way in which we introduced fluctuating boundary conditions to enforce a zero total current in the $\mathcal{J}$ variables. In particular,  these variables appear as $\nabla_\mu\phi$ everywhere in the system, except at the cuts where we added the $\gamma$ variables, where they appear as $\nabla_\mu\phi-\gamma_\mu$. 
Assuming here that $r_\mu$ takes values $0, 1, \dots, L_\mu-1$ and the fluctuating boundary condition is between $r_\mu = L_\mu-1$ and $r_\mu = 0$, we can make a change of variables $\phi \rightarrow \tilde{\phi}$, 
\begin{equation}
\tilde{\phi}(r) = \phi(r) + \sum_\mu \gamma_\mu r_\mu/L_\mu ~.
\label{phitilde}
\end{equation}
This gives $\nabla_\mu \phi(r) - \gamma_\mu \delta_{r_\mu = L_\mu-1} = 
\nabla_\mu \tilde{\phi}(r) - \gamma_\mu/L_\mu$, and the action becomes translationally invariant along the edge in the $\tilde{\phi}$ variables, so we will use them to measure correlators.

In the $\tilde{\phi}_1$ variables, we measure the correlator
\begin{equation}
\chi_1(r-r')\equiv\langle e^{i\tilde{\phi}_1(r)}e^{-i\tilde{\phi}_1(r')}\rangle.
\label{Crr2}
\end{equation}
Gapless modes exist when $\chi_1(r-r')$ has algebraic decay along the edge. We have taken $r'=r+m\hat{\mu}$, where $m$ is an integer such that $0<m<L$ and $\hat{\mu}$ is a unit vector in either the $y$ or $z$ directions. When we present numerical data we have averaged over all directions and all sites on the edge. We can choose either edge to measure the correlators at; in this work we have measured at $X=X_R$.  Since the edge is defined with respect to the dual lattice, it is not obvious where on the direct lattice to perform measurements of the edge states.  If the sites of the direct lattice are indexed by $(x,y,z)$, in our numerics we defined the sites of the dual lattice to be located at $(x+1/2,y+1/2,z+1/2)$. Using this definition and Eq.~(\ref{Sfrac1}) we can see that the edge effects will be most noticeable at $x=X_L-1/2$ or $x=X_R-1/2$, and so we measured the $\phi_1$ correlators at $x=X_R-1/2$. 

We can gain some insight into the behavior of these correlators by comparing them to spin-wave theory with the action
\begin{equation*}
S_{\text{edge, spin-wave}}[\phi_1]=\int dydz \frac{\lambda_1}{2}\left[ (\nabla_y \phi_1)^2+(\nabla_z \phi_1)^2 \right].
\end{equation*}
This is a good approximation to Eq.~(\ref{Sfrac1}) in the limit of small $\lambda_1$ and $\lambda_2$, where vortices in the $\phi_1$ variables (which are equivalent to $\mathcal{Q}_1$ variables) have a very large energy cost to cross the edge. Indeed, in the limit $\lambda_2\rightarrow0$ one can see that no vortices $\mathcal{Q}_1$ are crossing the plane where we measure the correlators, and the spin wave theory is appropriate on this plane. The spin-wave theory predicts the algebraic decay exponent of $\chi_1$ to be 
\begin{equation}
b_{\chi_1} \approx 1/(2\pi\lambda_1).
\label{bchi1_sw}
\end{equation}
We expect our data to be consistent with this prediction at small $\lambda_1$ and $\lambda_2$. At larger $\lambda_2$ there will be more vortices $\cQ_1$ in our system which increase the decay exponent of the $\phi_1$ correlations, while at larger $\lambda_1$ in the bulk the correlations can also develop through the bulk terms. In our numerics taken at $\lambda_1 = \lambda_2$, the first effect dominates and the extracted power law exponent is larger than this spin wave prediction.

To detect gapless modes in the $\mathcal{J}_2$ variables, we measure the correlator:
\begin{equation}
\chi_2(R-R')\equiv\langle e^{ic\tilde{\phi}_2(R)}e^{-ic\tilde{\phi}_2(R')}\rangle.
\label{Crr}
\end{equation}
The $c$ in Eq.~(\ref{Crr}) is the same as the one defining the quantum Hall state with $\sigma^{12}_{xy}=2c/d$.  As discussed after Eq.~(\ref{Sfrac2}), in the limit where $\lambda_2$ is very small we can change a $\phi_2$ variable by $2\pi/c$ while also redefining $\vec{p}_2$ and $\JJ_1$ variables and incur only a very small energy cost. Therefore for $c>1$ if we only measured correlators of $e^{i\phi_2(R)}$ we would not see any order since each $\phi_2$ variable will be randomly distributed in one of $c$ orientations, and it is the fluctuations around these orientations that will be power-law correlated.  Thus, our direct analysis suggests that for $c>1$ only $c$-tupled ``molecular'' states of type-2 bosons can propagate along the edge.

The $\chi_2$ data can be compared to the following ``spin-wave'' reasoning. Consider for simplicity edges of the $\sigma^{12}_{xy} = 2$ quantum Hall state and reformulation Eq.~(\ref{Sint}). In this case we can easily see the mechanism which forbids vortices on the edge. In the limit of small $\lambda_1$, we expect the currents of $\vec{M}$ to be zero everywhere away from the edges, while they satisfy a conservation law on the edge, $\sum_{\mu=y,z} \nabla_\mu M_\mu = 0$.  The action on the edge at $X=X_R$ has the structure 
\begin{equation*}
S_{\rm edge}[\phi_2, M_y, M_z] = \frac{1}{2\lambda_1 (2\pi)^2} \sum_{R \in \text{edge}; \mu=y,z}[\nabla_\mu \phi_2 - 2\pi P_\mu]^2,
\end{equation*}
with $P_y \equiv M_z$ and $P_z \equiv -M_y$.  For such a 2D XY model, the curl of the field $(P_y, P_z)$ is the vorticity, but this curl is precisely $\sum_{\mu=y,z} \nabla_\mu M_\mu$, which is zero.  Thus, in this limit, the action is like an XY model with completely prohibited vortices.  Away from this limit, we expect the vortices to still be effectively prohibited, and the spin wave treatment is now justified.  From examining Eqs.~(\ref{Sint}) and (\ref{Sfrac2}), we see that at small $\lambda_1$ and $\lambda_2$ such spin-wave theory predicts the exponent 
\begin{equation}
b_{\chi_2} \approx 2\pi\lambda_1 d^2
\end{equation}
for the algebraic decay of $\chi_2(R-R')$.  

We now present results for $\chi_1$ and $\chi_2$. In all the results in this section, we show data on the line in the parameter space where $\lambda_1=\lambda_2=\lambda$. All data was taken with a system size of $L=20$.  We know that we are in the quantum Hall phase when $\lambda$ is small, because it is here that the $\mathcal{G}$ variables are gapped. In Sec.~\ref{sec:reverse} we will present the phase diagrams of these models and will see that, for all $\eta \neq 0$, the system is in this quantum Hall state for $\lambda d^2\lesssim 0.33$ (the precise value depends on $c$ and $d$, but not very sensitively), while for $\eta=0$ the system will be in a trivial insulator state in this parameter regime.  Therefore at small $\lambda$ the edge we are studying is between a trivial insulator and a quantum Hall insulator. We have measured the correlation functions for $\lambda d^2=0.07,0.1,0.15,0.2,0.25$ and seen algebraic decay.  [We quote $\lambda d^2$ because at larger $d$ it takes smaller values of $\lambda$ for the $\cal{G}$ variables to become gapped, as can be seen in Eq.~(\ref{vgd2}).]  We plot $\chi$ on a log-log plot, so that if it decays algebraically we will see straight lines.  Due to the finite size of the system the lines will not be perfectly straight. We compensate for this by replacing $R-R'$ with the `chord distance' $D_{RR'}$:
\begin{equation}
D_{RR'}=\frac{L}{\pi}\sin\left[\frac{\pi |R-R'|}{L}\right],
\label{cordlength}
\end{equation}
which is often used in studies of (1+1)D systems.
With this substitution the plots should exhibit straight lines, and we found that this works very well in the present cases.

\begin{figure}
\includegraphics[width=\linewidth]{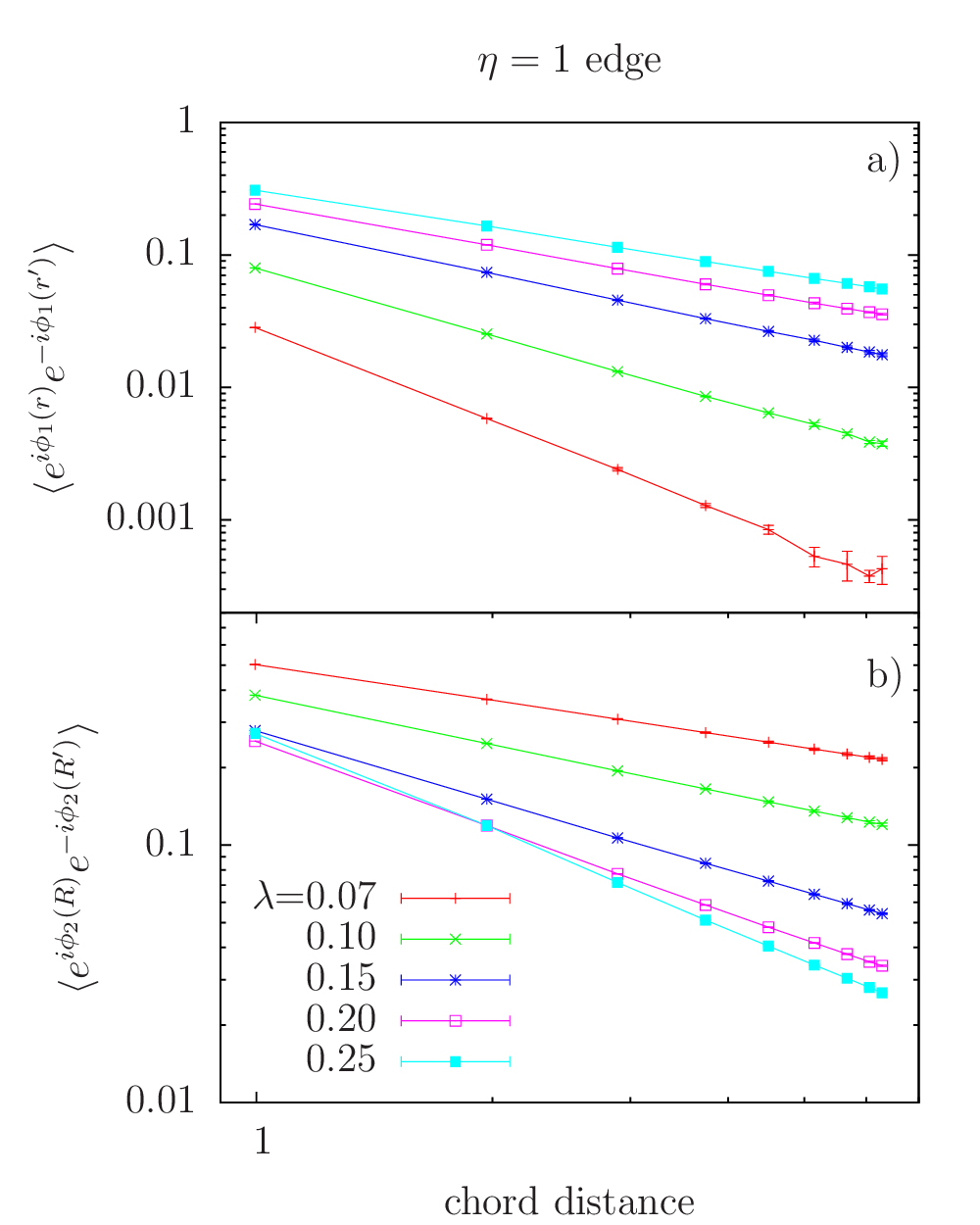}
\caption{ Correlation functions a) $\chi_1$ and b) $\chi_2$, plotted against the chord distance of Eq.~(\ref{cordlength}), on a log-log scale, for $\eta=1$ edge.  Error bars come from comparing runs with different initial conditions.  The straight lines imply that we have algebraic decay in the correlation functions and therefore the edge is gapless.  The slope of these lines varies with $\lambda$, and the extracted exponents are shown in Fig.~\ref{exponents}.}
\label{onegood}
\end{figure}

In Fig.~\ref{onegood} we show $\chi_1$ and $\chi_2$ for the case where $\eta=1$. The plot contains straight lines, which means that we do have algebraic decay. We can see that the slope of these lines, and therefore the decay exponents, depend on $\lambda$. In taking the data for this plot we used the reformulation given by Eq.~(\ref{Sint}) for the $\chi_2$ measurements. This reformulation is more efficient than the one in Eq.~(\ref{Sfrac2}), and allows us to obtain better statistics. However, we have compared the results of the two reformulations and found them to be consistent. Note that as $\lambda$ is increased, $\chi_1$ decays more slowly while $\chi_2$ decays more quickly. This behavior suggests that $\phi_1$ and $\phi_2$ behave as conjugate variables, which is one of the predictions of the $K$-matrix theory discussed in \ref{app:connections}.

\begin{figure}
\includegraphics[width=\linewidth]{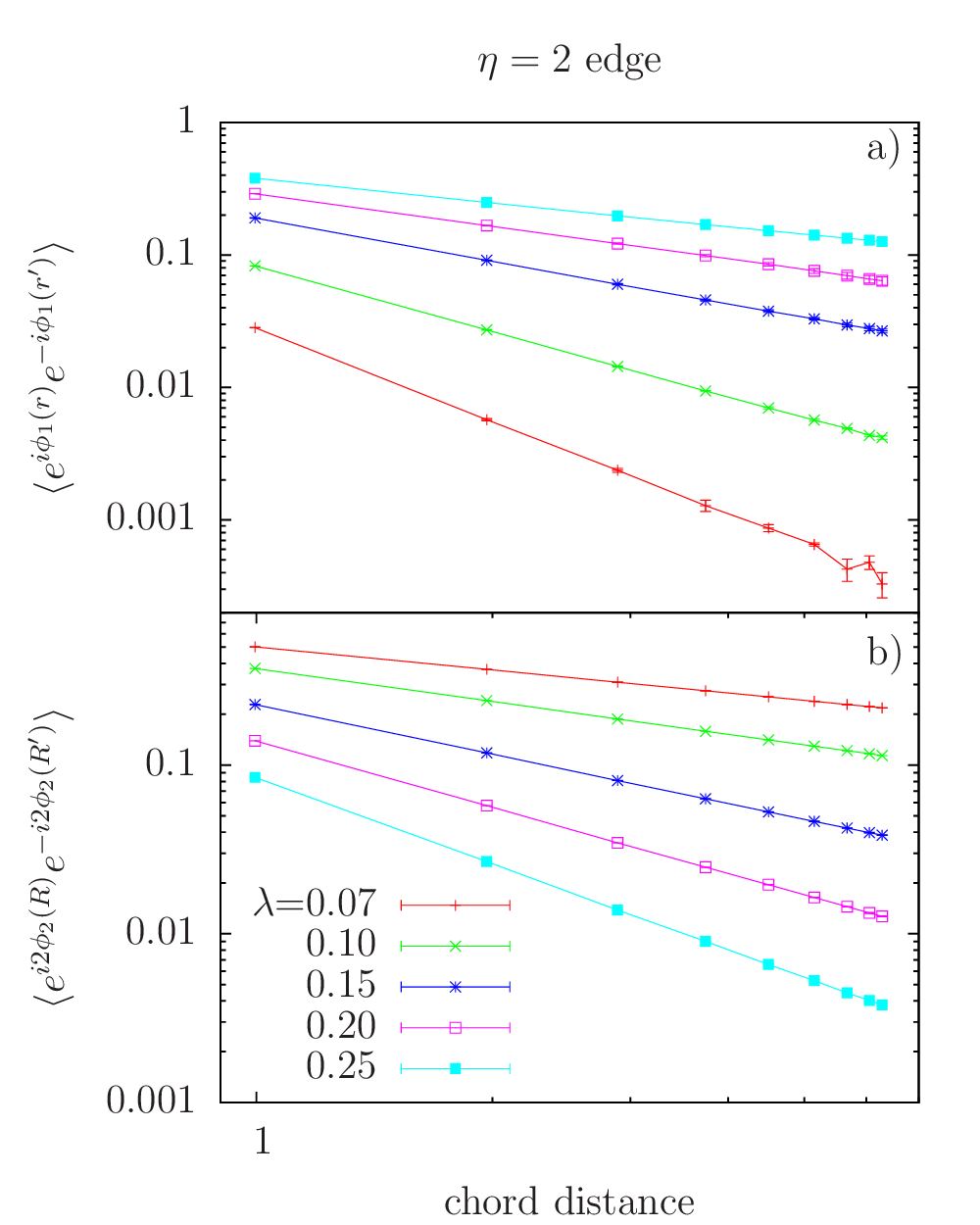}
\caption{ Same as Fig.~\ref{onegood}, but for $\eta=2$ edge. Once again, we see evidence for gapless modes. Note that in b) we measure pair-boson $e^{i 2\phi_2}$ correlators on the edge, while single-boson $e^{i\phi_2}$ correlators decay exponentially.
\label{twogood}}
\end{figure}

Figure \ref{twogood} shows the same measurements for $\eta=2$. Again we see evidence of algebraic decay with exponents which depend on $\lambda$. As above, we used the reformulation in Eq.~(\ref{Sint}) for the $\chi_2$ measurements. Note that for this data it is important to measure correlators of $\exp(i2\phi_2)$, as defined in Eq.~(\ref{Crr}). As expected, we found that single-boson correlators decay exponentially in this case, and only pair-boson correlators show algebraic decay.

\begin{figure}
\includegraphics[width=0.6\linewidth,angle=-90]{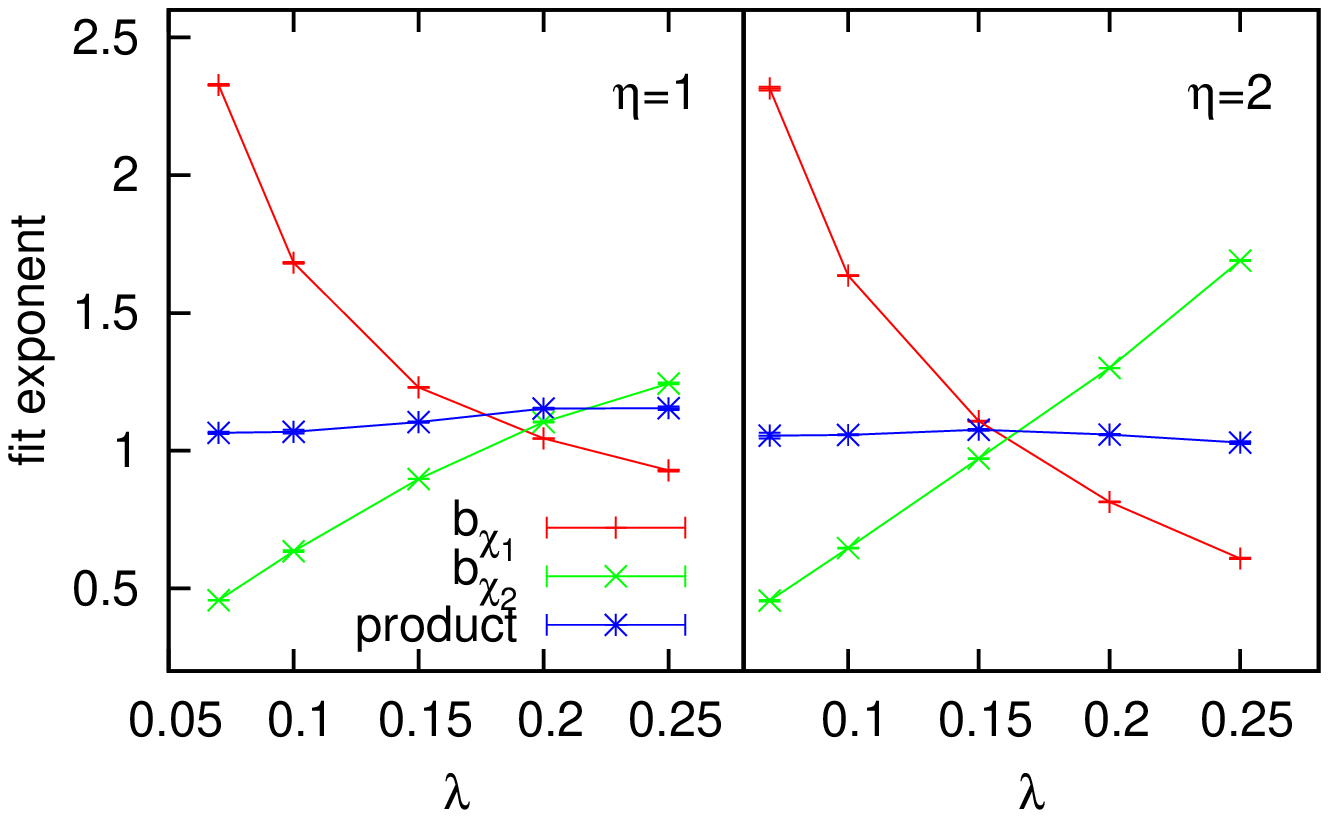}
\caption{ The exponents of the algebraic decay for $\eta=1$ and $\eta=2$, extracted using the fitting function in Eq.~(\ref{fitfunction}). We see that the $b_{\chi_1}$ exponent decreases with increasing $\lambda$ while the $b_{\chi_2}$ exponent increases. The product of these exponents is also shown.
\label{exponents}}
\end{figure}

From the above data we can extract the exponents of the algebraic decay. We fit the above data to the function
\begin{equation}
\chi_1(r-r')=\frac{A}{\left[\frac{L}{\pi}\sin\left(\frac{\pi|r-r'|}{L}\right)\right]^{b_{\chi_1}}},
\label{fitfunction}
\end{equation}
with $A$ and $b_{\chi_1}$ parameters of the fit. We analyzed $\chi_2$ similarly.  Figure~\ref{exponents} shows plots of these exponents for $\eta=1$ and $\eta=2$. The decay exponents for both $\chi_1$ and $\chi_2$ are slightly above the spin-wave prediction at small $\lambda$ (e.g., within 10\% for $\lambda =0.07$). At large $\lambda$ the fitted exponents differ significantly from the naive spin-wave predictions, though within an order of magnitude. In \ref{app:connections} we discuss a phenomenological understanding of the edge which predicts that the product of these exponents in the integer quantum Hall case should be equal to $1$. We can see from Fig.~\ref{exponents} that the products we measured are approximately equal to one.

\begin{figure}
\begin{center}
\includegraphics[width=0.7\linewidth]{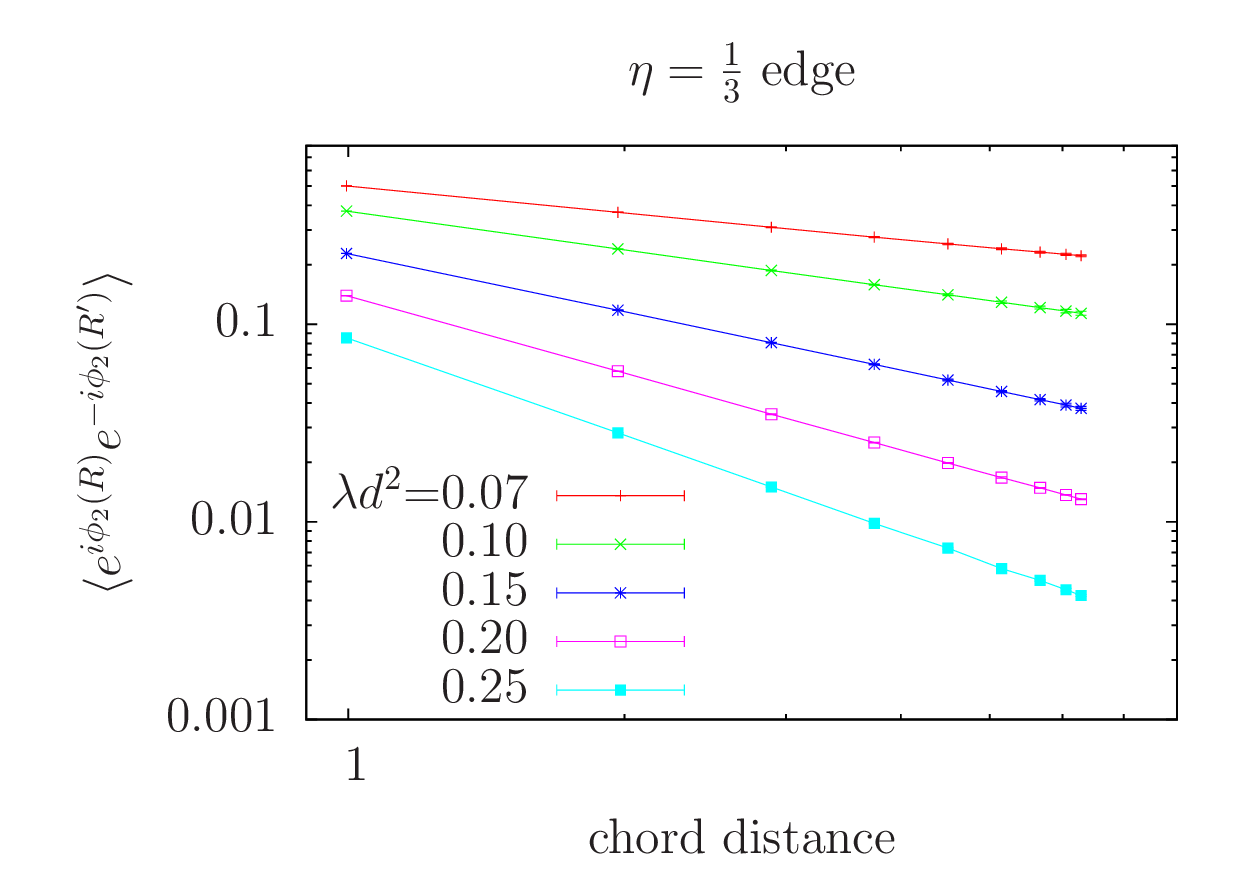}
\caption{ $\chi_2$ correlation function for $\eta=1/3$ edge, where $c=1$, $d=3$. The straight lines indicate algebraic decay for this fractional case. We were unable to obtain data for $\chi_1$ because it decayed too quickly.
\label{onethird}}
\end{center}
\end{figure}

Figure~\ref{onethird} shows $\chi_2$ for $\eta=1/3$. The existence of straight lines in this plot implies that we have gapless modes in a fractional quantum Hall system. We acquired this data using the reformulation in Eq.~(\ref{Sfrac2}). We were unable to measure $\chi_1$ for the fractional cases because the decay exponents were too large. Recall that in order to have gapped $\cal{G}$ variables we need $\lambda d^2 \lesssim 0.33$; for $d^2=9$ this leads to a small $\lambda$, and the spin-wave theory estimate Eq.~(\ref{bchi1_sw}) tells us that this leads to large exponents for $\chi_1$.  The presence of the factor $d^2$ in the product $b_{\chi_1} b_{\chi_2} \approx d^2$, which we obtained here by the direct analysis, is an indirect manifestation of the fractionalization when parameter $d>1$.  Indeed, in \ref{app:connections} we use a phenomenological model of the edge and the fractionalization of the particles to show that the product of these exponents should be equal to $d^2$, while if there is no fractionalization the product will be equal to $1$. In our lattice model of the edge and the specific parameterizations of the potentials, we have found that the values of $b_{\chi_2}$ are numerically similar to those in the integer case, but the $b_{\chi_1}$ values are much larger. This implies that the product of the exponents is greater than $1$, providing indirect evidence for fractionalization.

%%%%%%%%%%%%%%%%%%%%%%%%%%%%%%%%%%%%%%%%%%%%%%%%%%%%%%%%%%%%%%%%%%%%%%%%
\section{Phase diagrams}
\label{sec:reverse}

In previous works\cite{Loopy,short_range3,Gen2Loops} we have studied actions similar to that in Eq.~(\ref{kaction}). However, in those works we considered only the case where $\theta(k)$ was equal to a rational constant multiplied by $2\pi$. These are precisely the actions of the $\mathcal{G}$ variables which appeared in Eq.~(\ref{gaction}), and from now on we will refer to them as ``statistical'' actions. The statistical variables are quasiparticles in the quantum Hall phases.  In Refs.~\cite{Loopy, short_range3} we did not attempt to connect the statistical actions to a physical system, and instead focused on their phase diagrams.  Therefore we did not specify the ``vacuum'' of physical variables as it does not affect the dynamics of the phase transitions, but it is this vacuum which carries the quantized $\sigma^{12}_{xy}$ as can be seen from Eq.~(\ref{gaction}).  Furthermore, we can now also specify physical charges of the quasiparticles.  

If we start with a statistical action with rational $\theta_{\cG}$, we can invert the change of variables procedure in Sec.~(\ref{sec::demon}) with the correct choice of $a,b,c,d$ to get an action with $\theta(k)\sim k^2$, as in Eq.~(\ref{action}), which we will from now on call the ``physical'' action. This is how we obtained the specific potentials in Eqs.~(\ref{vintro}) and (\ref{tintro}). We know from the previous works the phase diagrams of the statistical actions in terms of the variables $\mathcal{G}$. We can use the change of variables in this work to describe these phase diagrams in terms of the physical variables $\mathcal{J}$. 

Recall that the specific potentials can be classified by the coefficients $c$ and $d$. When changing from the physical variables $\mathcal{J}$ to the statistical variables $\mathcal{G}$, we need these coefficients as well as $a$ and $b$, but these coefficients are not independent since they must satisfy $ad-bc=1$. In particular, if we have one solution $a_0$, $b_0$ then this constraint tells us that 
\begin{eqnarray}
&&a=a_0 + mc,\nonumber\\
&&b=b_0 + md,
\label{bshift} 
\end{eqnarray} 
are also solutions if $m$ is an integer.  The statistical actions can be classified by their statistical angle, which for our specific potential choices is given by $\theta_\cG=2\pi b/d$. Therefore each physical action can be related to multiple statistical actions by our change of variables. However, these statistical actions differ only in that $\theta_\cG$ can be different by an integer multiple of $2\pi$. Such a shift will have no effect on the partition sum in the $\mathcal{G}$ variables. Therefore all of the statistical actions which can be related to a given physical action have the same behavior. 

We can also see that multiple physical actions whose ratio $c/d$ differ by an integer can also be mapped to the same statistical action: that is, the actions in terms of $\cal{G}$ particles are essentially the same except for the ``background'' quantum Hall conductivity $\sigma^{12}_{xy}$ changing by an even integer. These physical actions are related to each other by adding an ``integer quantum Hall layer'' to the system without changing the properties of the fractionalized excitations. 

To begin the discussion of the broader phase diagrams of our models, it is useful to consider also the action in terms of the dual variables $\mathcal{Q}_1$, $\mathcal{Q}_2$:
\begin{eqnarray}
S &=& \frac{1}{2}\sum_k \frac{(2\pi)^2 }{|\vec{f}_k|^2} \left[\lambda_1|\QQ_1(k)|^2+\lambda_2|\QQ_2(k)|^2\right]
+ i\sum_k \frac{2\pi c}{d} \QQ_1(-k)\cdot \vec{a}_{\mathcal{Q}2}(k),
\label{qaction}
\end{eqnarray}
where $\QQ_2=\curl\vec{a}_{\mathcal{Q}2}$. This action comes from dualizing the $\cJ_1$, $\cJ_2$ variables with the specific potentials in Eqs.~(\ref{vintro}) and (\ref{tintro}). [Eq.~(15) in Ref.~\cite{short_range3} contains this action with general potentials.] This action can also be obtained by applying the modular transformation $(0, -1, 1, 0)$ to the original action. Note that the $d$ in Eq.~(\ref{qaction}) corresponds to the parameter in Eqs.~(\ref{vintro})-(\ref{tintro}), and is not related to the modular transformation used to obtain this phase.  The $\cQ$ variables are vortices with the usual long-range intra-species interactions, $v_{\cQ}(k) \sim 1/k^2$ in momentum space, and we see that these vortices are gapped for large $\lambda_1$ and $\lambda_2$.  The statistical angle in the third term of Eq.~(\ref{qaction}) is a rational number.  However, this rational number is unique to the specific choices we made in Eqs.~(\ref{vintro})-(\ref{tintro}), and small short-range modifications of the model can lead to a different statistical angle for the $\mathcal{Q}$ variables.  This is unlike the rational $\theta_{\mathcal{G}}$ in the quantum Hall insulators which is robust to short-range modifications of the potentials.  The difference comes from the qualitative difference when applying Eq.~(\ref{VJ}) to generic short-ranged $v(k) \sim {\rm const}$ and $\theta(k) \sim k^2$ in the cases $d=0$ (duality to only vortices) and $d\neq 0$ (more general modular transformation).

\subsection{Models with $c/d=n$, $\sigma^{12}_{xy}=2n$}
\begin{figure}
\begin{center}
\includegraphics[width=0.7\linewidth]{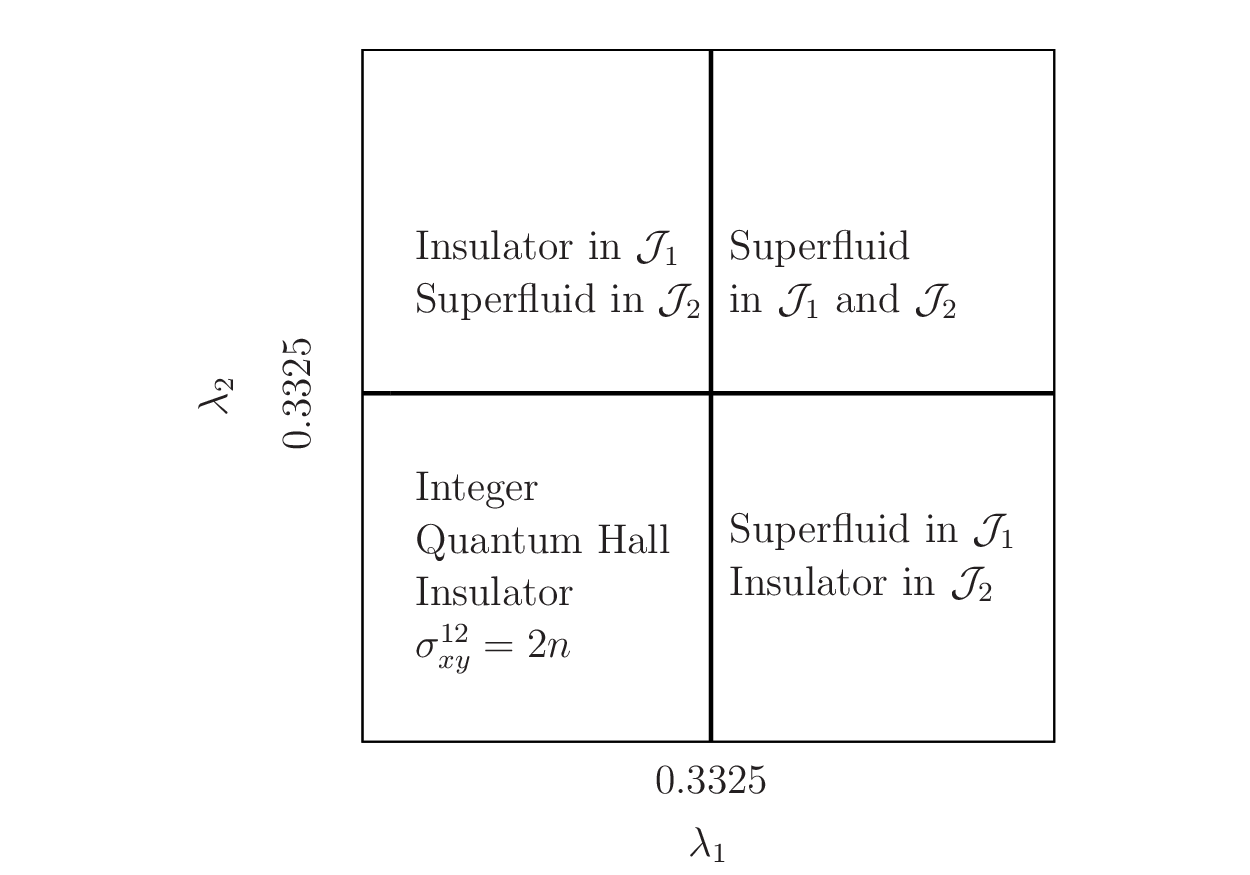}
\caption{The phase diagram for the model with the potentials of Eqs.~(\ref{vintro})-(\ref{tintro}) and $c=n$, $d=1$. In the lower left phase the $\mathcal{G}$ variables are gapped and we have the integer quantum Hall phase with $\sigma^{12}_{xy}=2n$. In the upper right phase the $\mathcal{J}$ variables are condensed, and we have a superfluid.  In the off-diagonal phases, one of the $\cJ$ variables is condensed and the other is gapped.}
\label{intphase}
\end{center}
\end{figure}

We now turn to detailed descriptions of the phase diagrams.
First we discuss the case where the conductivity in the quantum Hall phase is quantized as an even integer. In this case we start with a physical action with the potentials Eqs.~(\ref{vintro})-(\ref{tintro}) with parameters $c=n$, $d=1$ for $n$ an integer. We can get a statistical action by applying the modular transformation $(1, 0, n, 1)$, and this gives a statistical action with $\theta_{\mathcal{G}}=0$ and a background Hall conductivity of $\sigma^{12}_{xy}=2n$. Therefore in the statistical action we have a system of  two uncoupled loops, which is a system that is well understood.\cite{Cha1991,Sorensen} 
In this system when $\lambda_i\lesssim 0.3325$, the variables $\mathcal{G}_i$ are gapped, and when $\lambda_i$ is greater than this value the $\mathcal{G}_i$ are condensed. All phase transitions are second-order XY transitions. Figure \ref{intphase} shows this phase diagram. 
Since we know the behavior of the $\mathcal{G}$ variables everywhere in the phase diagram, we can now deduce the behavior of the physical $\mathcal{J}$ variables. In the lower left corner phase we have seen that the $\mathcal{J}$ variables are in a quantum Hall phase. 

To understand the rest of the phase diagram we must make more precise our earlier definitions of what makes a variable ``gapped'' or ``condensed''. When a variable is in a phase in which it is gapped, the energy cost for having large loops of that variable becomes arbitrarily large and only small loops are present. When a variable is condensed the energy cost for forming loops is small. A variable is condensed if and only if the variable dual to it is gapped. In some phases a variable will be neither condensed nor gapped in the above sense; instead the variable can be part of a composite object which is condensed or gapped. This is the situation for the $\cJ$ variables in the quantum Hall phases. We bring this out to indicate that there are more cases than just given by binary choice of $\cJ_i$ variable being gapped or condensed.  In all cases, the precise meaning is provided by finding appropriate transformation that leads to a description in terms of gapped particles only.

With this in mind, we can interpret the rest of the phase diagram. We begin with the phase in the upper-right corner, where $\lambda_1$ and $\lambda_2$ are large. We can see that in this phase the potentials seen by the $\mathcal{Q}$ variables in Eq.~(\ref{qaction}) become arbitrarily large, so both of these species of variable are gapped. Therefore both species of $\mathcal{J}$ variable are condensed and this phase is a superfluid. The conductivities $\sigma^{11}_{xx}$ and $\sigma^{22}_{xx}$ diverge in this phase, while the Hall conductivity $\sigma^{12}_{xy}$ is non-universal. 

We now study the off-diagonal phases in Fig.~\ref{intphase}. For simplicity we discuss the phase in the lower right corner where $\lambda_1$ is large but $\lambda_2$ is small (in fact $\lambda_1$ can become arbitrarily large in this phase). The upper left corner is similar with the indices interchanged. From Eq.~(\ref{vintro}) we can see that when $\lambda_1\rightarrow\infty$,  $v_2(k)\rightarrow\frac{1}{\lambda_2}$. Since $\lambda_2$ is small in this phase, $v_2(k)$ can become arbitrarily large and the $\mathcal{J}_2$ variables must be gapped. In addition, we can see from Eq.~(\ref{JQreal}) that the $\mathcal{Q}_1$ variables see an arbitrarily large potential and are therefore gapped, so the $\mathcal{J}_1$ variables are condensed. From the above results, we can conclude that this phase is a trivial insulator in the $\mathcal{J}_2$ variables and a superfluid in the $\mathcal{J}_1$ variables.

Finally, we note that the specific model, Eqs.~(\ref{vintro})-(\ref{tintro}), with $c=n, d=1$ does not realize the trivial insulator phase with both $\cJ_1$ and $\cJ_2$ gapped.  Of course, we can obtain such a phase by different modifications of the potentials, e.g., by adding large repulsive pieces to both $v_1$ and $v_2$, and it would be interesting to study such models in the future.

%%%%%%%%%%%%%%%%%%%%%%%%%%%%%%%%%%%%%%%%%%%%%%%%%%%%%%%%%%%%%%%%%%%%%%%%
\subsection{Models with $d\neq 1$}
\begin{figure}
\begin{center}
\includegraphics[width=0.7\linewidth]{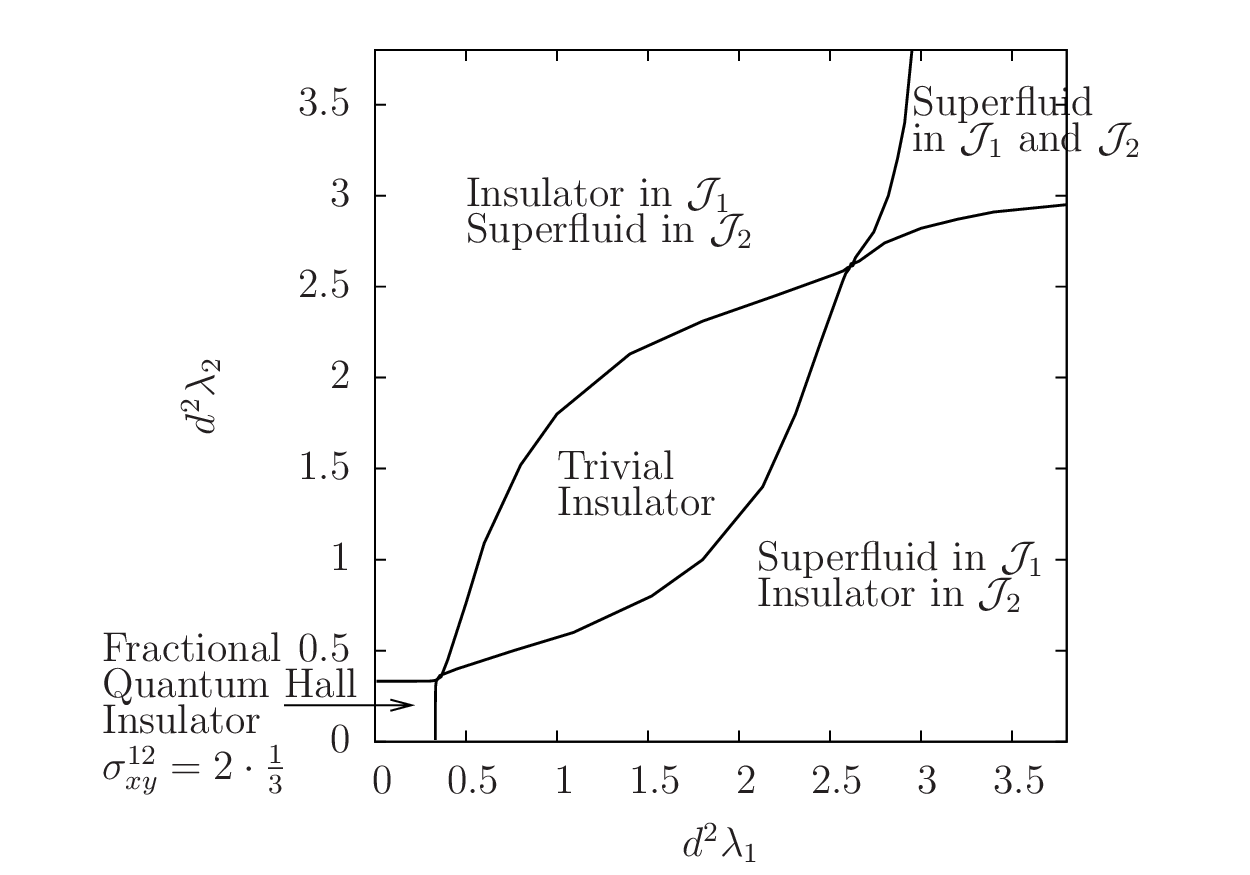}
\caption{The phase diagram for the model with $c=1,d=3$. In the lower left phase the $\mathcal{G}$ variables are gapped and we have a fractional quantum Hall phase with $\sigma^{12}_{xy}=2\cdot\frac{1}{3}$. In the upper right phase the $\mathcal{J}$ variables are condensed, which implies a superfluid. In the middle phase the $\mathcal{J}$ variables are gapped and we have a trivial insulator. This figure is reproduced from Ref.~\cite{short_range3}, but the phases have been re-interpreted in terms of the physical variables discussed in this work.}
\label{fracphase}
\end{center}
\end{figure}

In Ref.~\cite{short_range3} we studied a statistical action with $\theta_{\cal{G}} = 2\pi/3$. The phase diagram for this model is shown in Fig.~\ref{fracphase}. In the lower left corner the $\mathcal{G}$ variables are gapped and this is the fractional quantum Hall phase. Any physical action with $d=3$ and $c=1+3m$ (for $m$ an integer) can be related to this statistical action by our modular transformation; here we will discuss the case where $c=1$. In this case the change of variables needed to get from the $\mathcal{J}$ variables to the $\mathcal{G}$ variables is $(0, -1, 1, 3)$. The fractional quantum Hall phase will have $\sigma^{12}_{xy}=2\cdot\frac{1}{3}$ and excitations carrying respective fractional charges of $1/3$ and mutual statistics of $2\pi/3$.

We know from our previous numerical study\cite{short_range3} that in the middle phase the variables dual to the $\mathcal{G}$ variables are gapped. 
We can also compute the action for the variables dual to the $\mathcal{G}$ variables and see that it has the same potential as the action for the $\mathcal{J}$ variables. The two actions also have values of $\theta(k)$ which differ only by an integer multiple of $2\pi$. Such difference will translate to factors $e^{2\pi i}$ in the partition sum and therefore will not contribute. Therefore if the variables dual to the $\mathcal{G}$ variables are gapped then the $\mathcal{J}$ variables should also be gapped. Therefore this middle phase is a trivial insulator in the $\mathcal{J}$ variables. This can be confirmed by measuring the conductivity numerically in this phase. 

In the upper right corner phase we can see from Eq.~(\ref{qaction}) that the $\cQ$ variables are gapped,  and therefore the $\mathcal{J}$ variables are condensed and this phase is a superfluid.  In our previous work\cite{short_range3} we found that transition between the trivial insulator and the superfluid is a pair of XY transitions, while we found more complicated behavior at the fractional quantum Hall-trivial insulator transition. 

The structure described in the previous paragraph holds for any set of physical variables which can be mapped to a statistical action with $\theta_\cG=2\pi/m$, with $m$ an integer. %\cite{footnoten2}.
One exception is $m=2$, where our specific model with $c=1, d=2$ has an additional symmetry in the $\cG$ variables which prevents the existence of the middle phase, cf.~Fig.~1 in Ref.~\cite{Loopy}.  This is discussed in detail in Refs.~\cite{Loopy, Gen2Loops}, while here we note that generic perturbations to our original model will break this symmetry and open a sliver of the trivial phase in the phase diagram.

Finally, for more complicated fractions $c/d$, our model will have multiple phases in the middle of the phase diagram, resembling hierarchy of phases that we found in $U(1) \times U(1)$ loop models with marginally long-ranged interactions and modular invariance.\cite{Gen2Loops}  We expect that the ``middle'' phase at the largest $\lambda$ is a trivial insulator, while the other phases are various quantum Hall states.  For example, in our model Eqs.~(\ref{vintro})-(\ref{tintro}) with $c=2, d=5$, we found the following sequence of phases upon increasing $\lambda_1 = \lambda_2$: fractional quantum Hall insulators $\sigma^{12}_{xy} = 2\cdot 2/5$ and $\sigma^{12}_{xy} = 2\cdot 1/2$, trivial insulator $\sigma^{12}_{xy} = 0$, and superfluid.  It would be interesting to explore such phase diagrams and phase transitions in more detail in the future.

%%%%%%%%%%%%%%%%%%%%%%%%%%%%%%%%%%%%%%%%%%%%%%%%%%%%%%%%%%%%%%%%%%%%%%%%
\section{Discussion}
In this work we have presented physical $U(1) \times U(1)$ bosonic models which realize insulating phases with a quantized Hall conductivity that can take both integer and fractional values. 
In the fractional case, we also have excitations carrying fractionalized charges and non-trivial mutual statistics. We have shown how to study these models in Monte Carlo and found evidence for gapless edge modes. We have also presented broader phase diagrams of our models.

The action in Eq.~(\ref{action}) can be derived from a local Hamiltonian, as shown in \ref{app:H}. When we included an edge in our action by varying $\eta(R)$ in Eq.~(\ref{JQreal}), we do not know precisely how that edge is realized in the physical Hamiltonian. It is possible that this method of including an edge changes the Hamiltonian near the edge in such a way as to create gapless modes which are not due to the bulk topological state of the system on one side.  For example, a local strengthening of boson hopping along the edge could lead to gapless (1+1)D Luttinger liquid modes.  Irrespective of the microscopic details, the edges that we studied do produce the quantum Hall $\sigma^{12}_{xy}$, so at least some of the observed properties are due to the topology of the bulk phases.  Including edges using different methods, and confirming that the observed gapless properties are not artifacts of the method used in this work, is a possible subject of future research. There is however some evidence that our gapless modes are due to topological effects. In the $\eta=1/3$ case we expect topological gapless modes to exist on the edge in the bottom left corner of the phase diagram, but not in the middle phase. This is precisely the behavior which we have observed, with gapless modes disappearing beyond $\lambda d^2=0.35$. In addition, we have observed gaplessness in both $\phi_1$ and $\phi_2$ variables (for the integer case where both signals could be detected), which is what we expect if the gapless modes are topological, and in qualitative agreement with the phenomenological $K$-matrix theory of the edge in ~\ref{app:connections}. 

Our work allows the numerical study of interacting topological insulator phases, and therefore may be able to address many questions about such phases. For example, we could investigate the effect of disorder and other perturbations on the gapless edge states. We could also study transitions between different topological phases. In Ref.~\cite{short_range3} we have observed unusual behavior at the fractional quantum Hall-trivial insulator transition, which could be studied more closely. In addition, our model can realize transitions between different fractional quantum Hall states, as well as transitions between integer quantum Hall states and trivial insulators, which are of recent interest.\cite{GroverVishwanath2012, LuLee2012_QPT}
More generally, it would be interesting to see what other interacting topological phases can allow unbiased numerical studies. Furthermore, since ideas in the present work do not rely on Chern-Simons construction specific to (2+1)D, they may be more readily extended to studies of such phases in higher dimensions.\cite{VishwanathSenthil2012, KeyserlingkBurnellSimon2013, XuSenthil2013, Wen2013}

\section*{Acknowledgements}
We would like to thank M.~P.~A.~Fisher, A.~Kitaev, T.~Senthil, and A.~Vishwanath, whose penetrating questions and suggestions inspired much of this work.  We also thank J.~Alicea and A.~Kapustin for useful discussions.  This research is supported by the National Science Foundation through grant DMR-1206096, and by the Caltech Institute of Quantum Information and Matter, an NSF Physics Frontiers Center with support of the Gordon and Betty Moore Foundation. SG is supported by an NSERC PGS fellowship. OIM would also like to acknowledge fruitful participation in the KITP program ``Exotic Phases of Frustrated Magnets''.

%%%%%%%%%%%%%%%%%%%%%%%%%%%%%%%%%%%%%%%%%%%%%%%%%%%%%%%%%%%%%%%%%%%
%%%%%%%%%%%%%%%%%%%%%%%%%%%%%%%%%%%%%%%%%%%%%%%%%%%%%%%%%%%%%%%%%%%
\appendix

\section{Formal duality procedure}
\label{app:duality}

In the main text we use a duality transform to make a change of variables, for example from the physical $\cJ$ variables to the dual $\cQ$ variables. In this appendix we review this duality explicitly for one loop species.\cite{PolyakovBook, Peskin1978, Dasgupta1981, FisherLee1989, LeeFisher1989, artphoton,short_range3}  The original degrees of freedom are conserved integer-valued currents $\JJ(r)$ residing on links of a simple 3D cubic lattice; $\vec{\nabla} \cdot \JJ(r) = 0$ for any $r$.  To be precise, we use periodic boundary conditions and also require vanishing total current, $\JJ_{\rm tot} \equiv \sum_r \JJ(r) = 0$.  We define duality mapping as an exact rewriting of the partition sum in terms of new integer-valued currents $\QQ(R)$ residing on links of a dual lattice and also satisfying $\vec{\nabla} \cdot \vec{\cQ}(R) = 0$ for any $R$ and $\vec{\cQ}_{\rm tot} = 0$. The derivation of this is as follows. As in Sec.~\ref{sec:reform} of the main text, we seek to integrate out the constrained, integer valued $\cJ$ variables.  We can implement these constraints using Eqs.~(\ref{zphi})-(\ref{zp}), obtaining the following partition sum:
\begin{equation}
Z=\int_{-\infty}^{+\infty}{\cal D}j_{\mu}(r) \sum_{p_{r\mu}=-\infty}^{+\infty}\int_{-\pi}^{\pi} {\cal D}\phi(r) \int_{-\pi}^{\pi} \prod_{\mu=1}^{3} d\gamma_{\mu} 
\exp\left(-S[j]+ i \sum_{r,\mu} j_\mu(r)[\nabla_\mu\phi(r) - 2\pi p_\mu(r)- \gamma_\mu \delta_{r_\mu=0}]\right),
\end{equation}
where $S[j]$ is the original action, now in terms of real valued variables. 

Now consider $\cQ$ variables such that $\QQ=\curl\vec{p}$. Clearly there are multiple values of $\vec{p}$ which give the same $\QQ$. Two such $\vec{p}$, given by $\vec{p}$ and $\vec{p}^0$, are related as follows:
\begin{equation}
p_\mu(r)=p^0_\mu(r)+\nabla_\mu N(r)+M_\mu \delta_{r_\mu=0}.
\end{equation}
Here $N(r)$ is an integer-valued field and $M_\mu$ are integers. We can divide all possible configurations of $p_\mu(r)$ into classes, where two configurations are in the same class if they can satisfy the above equation. We can separate the above sum over all $p_\mu(r)$ into a sum over classes (where each distinct class is denoted by a fixed member $p_\mu^0(r)$), as well as a sum over the members of each class, which corresponds to a sum over $N(r)$ and $M_\mu$. We can then absorb the sums over $N(r)$ and $M_\mu$ into the definitions of $\phi(r)$ and $\gamma_\mu$, which changes the limits on their integration to $(-\infty,+\infty)$. We can then interpret the integration over these variables as producing delta function constraints on the $j$ variables. This gives the partition sum:
\begin{equation}
Z=\int_{-\infty}^{+\infty}\prod_{\mu=1}^{3}{\cal D}j_{\mu}(r)\sum_{\QQ=\curl \vec{p}^0} \prod_{r\neq0} \delta[\divv\vec{j}(r)=0]\delta[\vec{j}_{\rm tot}=0]  
\exp\left[-S(j)- 2\pi i \sum_{r,\mu} j_\mu(r) p^0_\mu(r)\right].
\label{Z3}
\end{equation}
Note that due to the U(1) symmetry of the action we can fix $\phi(r=0)=0$, $N(r=0)=0$, and so there is no delta function at $r=0$. 

If we wish to obtain an action entirely in the $\cQ$ variables, we can now integrate out the $\vec{j}$ fields. In this work, we study actions of the form:
\begin{eqnarray}
S[\JJ]=\frac{1}{2}\sum_{k} v(k) |\JJ(k)|^2 + i\sum_k  \JJ(k)\cdot \vec{A}(-k),
\label{SJ}
\end{eqnarray}
where $v(k)$ is a potential and $\vec{A}$ is a fixed gauge field coupled to the $\JJ$ variables (in the main text in different contexts it corresponds to the external gauge field $\vAext$ or the internal gauge field $\vec{a}_{2}$).
With this action, the integrations over the $j$ variables are Gaussian, with basic averages with respect to the quadratic piece in Eq.~(\ref{SJ}) given by:
\begin{equation}
\la j_\mu(k) j_{\mu'}(k') \ra_0 = \frac{\delta_{k+k'=0}}{v(k)} \left(\delta_{\mu \mu'} - \frac{f_{k,\mu} f_{k,\mu'}^*}{|\vec{f}_k|^2} \right) ~,
\end{equation}
 where $f_{k,\mu} \equiv 1 - e^{i k_\mu}$.  We then obtain
\begin{equation}
S_{\rm dual}[\vec{Q}] = \frac{1}{2} \sum_k \frac{[2\pi\vec{Q}(-k) + \vec{B}(-k)] \cdot [2\pi\vec{Q}(k) + \vec{B}(k)]}{v(k) |\vec{f}_k|^2} ,
\label{SQ}
\end{equation}
where $\vec{B} \equiv \vec{\nabla} \times \vec{A}$.  The relation between Eq.~(\ref{SJ}) and Eq.~(\ref{SQ}) is what we call ``duality map'' in the main text.

We can also view the duality procedure in the following way, which is useful for the discussion of the K-matrix approaches in \ref{subsec:Kmatrix}. Starting with Eq.~(\ref{Z3}) we can enforce the constraints implemented by the delta functions by writing $j$ as the curl of a real valued field: $\vec{j}=(\curl\vec{a})/2\pi$. We can then perform unconstrained integration over all values of $a$, and the partition sum can be written as:
\begin{equation}
Z=\int_{-\infty}^{+\infty}\prod_{\mu=1}^{3}{\cal D}a_{\mu}(R)\sum_{\QQ} 
\exp\left[-S\left(\frac{\curl\vec{a}}{2\pi}\right)- i \sum_{R,\mu} a_\mu(R) \cQ_\mu(R)\right].
\end{equation}
We see that, compared to our original action $S[\cJ]$, duality corresponds to replacing integer valued $\JJ$ by real-valued $(\curl\vec{a})/2\pi$, and adding a term which couples the $\vec{a}$ fields to new variables $\QQ$.

\section{Relation to other approaches}
\label{app:connections}
In the main text, we obtained the physics of our models directly using exact transformations.  Here we point out connections to effective field-theoretic approaches that may be more familiar to the readers.  Our models can provide rigorous framing and testing grounds for such approaches.

%%%%%%%%%%%%%%%%%%%%%%%%%%%%%%%%%%%%%%%%%%%%%%%%%%%%%%%%%%%%%%%%%%%
\subsection{Relation to non-linear sigma models with topological terms}
Let us consider our model in terms of the dual vortex variables $\cQ_{1,2}$, which has an action given by Eq.~(\ref{qaction}).  As argued in Ref.~\cite{Senthil2006_theta}, a similar vortex loop model arises in an $O(2) \times O(2)$ theory with a topological term with $\theta = \theta_\cQ$.  We see that integer quantum Hall states in our model ($c=n, d=1$) correspond to $\theta_\cQ = 2\pi n$, as proposed in Ref.~\cite{SenthilLevin2012}.  We also see that fractional ``$c/d$'' Quantum Hall states constructed in this paper correspond to $\theta_\cQ = 2\pi c/d$.  Our models thus provide lattice regularizations of such non-linear sigma models with topological terms.  Here we want to emphasize that such models do not correspond to a single phase; instead, they can have rich phase diagrams, as illustrated in the main text.

Using our action in terms of the $\cQ_{1,2}$ variables [Eq.~(\ref{qaction})] we can obtain an equivalent formulation of the original physical current model as
\begin{eqnarray}
&&S[\valpha_{q1}, \valpha_{q2}, \JJ_1, \JJ_2] =
\frac{1}{2} \sum_k
\frac{\lambda_1 |[\vec{\nabla} \times \valpha_{q1}](k)|^2
      + \lambda_2 |[\vec{\nabla} \times \valpha_{q2}](k)|^2}{ |\vec{f}_k|^2} \\
&& ~~~~+ i\sum_k \frac{c}{2\pi d} [\vec{\nabla} \times \valpha_{q1}](-k) \cdot \valpha_{q2}(k) 
 + i \sum_k \left[\JJ_1(-k) \cdot \valpha_{q1}(k)
                    + \JJ_2(-k)\cdot \valpha_{q2}(k)\right] ~.
\nonumber
\end{eqnarray}
The above equation is an intermediate step in the exact duality transformation between $\QQ_{1,2}$ and $\JJ_{1,2}$, where $\valpha_{q1,2}$ are auxiliary real-valued gauge fields encoding conserved real-valued currents that appear in the duality transformation, see \ref{app:duality}.  We can interpret these gauge fields as mediating the interactions of the physical currents $\JJ_{1,2}$.  The action for the gauge fields has a mutual Chern-Simons term but also explicit ``mass terms''; the latter make the physical current interactions short-ranged, as desired.

Since the gauge fields are real-valued (non-compact), the above representation implies that our model action is unitary, i.e., it can arise as a path integral of a quantum Hamiltonian\cite{Gukov2004, Hansson2004} (see also direct demonstration in \ref{app:H}).  In fact, this is true for arbitrary formal parameter $c/d \to \eta$.  Note, however, that any such model can still have only rational Quantum Hall phases of the type described in the main text, with different rational $\sigma^{12}_{xy} = 2c'/d'$ in different regimes.  Indeed, the available transformations to new gapped variables are modular transformations with integer entries $(a', b', c', d')$ and can only produce gapped Quantum Hall phases with rational $\sigma^{12}_{xy}$.  We can still apply this method to analyze microscopic models with any $\eta$, where it is natural to try rational approximants $c'/d'$ to $\eta$ and hence natural to expect rich phase diagram,\cite{Cardy1982, Shapere1989, Gen2Loops} while details require case-by-case study.

Finally, we note close relation to one of the reformulations in the main text.  Since we ultimately want the action in terms of only the conserved currents $\JJ_1$ and $\JJ_2$, we can perform the integration over the gauge fields $\valpha_{q1}$ and $\valpha_{q2}$ in any gauge.  For example, we can use the gauge $\vec{\nabla} \cdot \valpha = 0$ and implement it as follows:  We replace $(\vec{\nabla} \times \valpha)^2 \to (\vec{\nabla} \times \valpha)^2 + \xi (\vec{\nabla} \cdot \valpha)^2$, perform unrestricted integration over the $\valpha$ variables, and take the limit of large $\xi$.  We can check that as long as the currents $\JJ$ are conserved everywhere, the unrestricted integration over $\valpha$ gives an action independent of $\xi$.  (A note of caution: the above statement does not hold when the currents have sources and sinks -- indeed, boson Green's functions are gauge-dependent.)  Taking specific value $\xi=1$ gives Eq.~(\ref{preal}) in the main text.
In the partition sum, we integrate independently over real-valued fields $\valpha_{1,2}$, and we can check directly that this gives the postulated model without the recourse to the dual description.    We reiterate that the action Eq.~(\ref{preal}) is not a gauge theory; rather, $\valpha_{1,2}$ are some local ``oscillator'' fields mediating short-ranged interactions of the physical currents $\JJ_{1,2}$.  In \ref{app:H}, we will show how such an action can arise as a path integral of a quantum Hamiltonian with only local degrees of freedom and local interactions.

%%%%%%%%%%%%%%%%%%%%%%%%%%%%%%%%%%%%%%%%%%%%%%%%%%%%%%%%%%%%%%%%%%%
\subsection{Relation to $K$-matrix theories}
\label{subsec:Kmatrix}
Here we show that our modular transformation analysis can be viewed as a derivation of a $K$-matrix-like theory, albeit with some non-standard structure in general.  The transformation consists of 
1) duality from $\JJ_1$ to $\QQ_1$;
2) change of variables $\QQ_1 = d \FF_1 - b \GG_2, \JJ_2 = c \FF_1 - a \GG_2$ [inverse of Eq.~(\ref{modularshift})];
and 3) duality from $\FF_1$ to $\GG_1$.
We show all steps starting with the physical action and give detailed explanations below:
\begin{eqnarray}
&& S[\JJ_1, \JJ_2; \vAext_1, \vAext_2] = S_{\rm s.r.}\left[\JJ_1, \JJ_2\right]
+ i \sum \left[\JJ_1 \cdot \vAext_1 + \JJ_2 \cdot \vAext_2 \right]; 
\label{SKmatr} \\
&& S_1[\vbeta, \QQ_1, \JJ_2; \vAext_1, \vAext_2] = S_{\rm s.r.}\left[\frac{\vec{\nabla} \times \vbeta}{2\pi}, \JJ_2 \right]
+ i \sum \left[\frac{\vec{\nabla} \times \vbeta}{2\pi} \cdot \vAext_1 + \JJ_2 \cdot \vAext_2 + \QQ_1 \cdot \vbeta \right]; \nonumber \\
&& S_2[\vbeta, \FF_1, \GG_2; \vAext_1, \vAext_2] = S_{\rm s.r.}\left[\frac{\vec{\nabla} \times \vbeta}{2\pi}, c \FF_1 - a \GG_2 \right] \nonumber\\
&& \quad ~+~ i \sum \left[\frac{\vec{\nabla} \times \vbeta}{2\pi} \cdot \vAext_1 + \FF_1 \cdot (d \vbeta + c \vAext_2) - \GG_2 \cdot (b \vbeta + a \vAext_2) \right]; \nonumber \\
&& S_3[\vbeta, \vgamma, \GG_1, \GG_2; \vAext_1, \vAext_2] = S_{\rm s.r.}\left[\frac{\vec{\nabla} \times \vbeta}{2\pi}, c \frac{\vec{\nabla} \times \vgamma}{2\pi} - a \GG_2 \right] + \nonumber \\
&& \quad ~+~ i \sum \left[d \frac{\vec{\nabla} \times \vgamma}{2\pi} \cdot \vbeta + \frac{\vec{\nabla} \times \vbeta}{2\pi} \cdot \vAext_1 + c \frac{\vec{\nabla} \times \vgamma}{2\pi} \cdot \vAext_2 \right]
+ i \sum \left[\GG_1 \cdot \vgamma - \GG_2 \cdot (b \vbeta + a \vAext_2) \right]. \nonumber
\end{eqnarray}
First, we do not need to specify the microscopic action $S_{\rm s.r.}$ with short-ranged interactions other than that it gives the desired ``$c/d$'' phase for some parameters; we carry $S_{\rm s.r.}$ throughout to better display all connections.  We also keep track of the external vector potentials $\vAext_{1,2}$.  At each step, we show explicitly degrees of freedom that form the partition sum.

1) We treat the duality from the boson current $\JJ_1$ to vortex current $\QQ_1$ as a reformulation of the partition sum replacing $\JJ_1 \to \vec{\nabla} \times \vbeta/(2\pi)$, with real-valued gauge field $\vbeta$, while keeping the information about the integer-valuedness of $\JJ_1$ with the help of new integer-valued current $\QQ_1$ as shown in $S_1$ (see, e.g.,~\ref{app:duality}).
2) Here we change to new independent currents $\FF_1, \GG_2$, which is a valid transformation for $(a, b, c, d)$ forming a modular matrix.
3) Finally, we perform formal duality from $\FF_1$ to $\GG_1$ as in 1): $\FF_1 \to \vec{\nabla} \times \vgamma/(2\pi)$ with real-valued gauge field $\vgamma$, plus new integer-valued current $\GG_1$, with the result in $S_3$.

As already mentioned, the $S_{\rm s.r.}$ part of the model is needed to stabilize the phase with gapped $\cG_{1,2}$ particles.  Once this is achieved, we can view $S_3$ as a $K$-matrix-like theory in terms of gauge fields $\vbeta, \vgamma$, with the matrix $K = \begin{pmatrix} 0 & d \\ d & 0 \end{pmatrix}$ and charge vectors $t_1^T = (1, 0)$ and $t_2^T = (0, c)$ for coupling to $\vAext_1$ and $\vAext_2$ respectively.  Because of the mutual Chern-Simons term for the gauge fields $\beta$ and $\gamma$, we can ignore $S_{\rm s.r.}$ and use standard $K$-matrix formalism\cite{Wen_book} to reproduce the result Eq.~(\ref{sigma}) in the main text for the $\sigma^{12}_{xy}$.  We can also reproduce the mutual statistics of the $\cG_1$ and $\cG_2$ quasiparticles Eq.~(\ref{tg}) and their charges Eq.~(\ref{charge1d}), but note that we must use the specific coupling of $\GG_2$ to $\vbeta$ and include the direct coupling to $\vAext_2$ (if $a \neq 0$) to obtain correct results.  For general $b$ and $a$, these are non-standard features of the theory in Eq.~(\ref{SKmatr}) compared to familiar $K$-matrix theories, but can be accommodated with proper care.
Of course, the recovery of all results is expected since the $K$-matrix formalism is simply gaussian integration over fields $\vbeta$ and $\vgamma$, while the transformations in the main text carry out such integrations implicitly but exactly for our model (also including the short-ranged piece $S_{\rm s.r.}$).
We remark that Eq.~(\ref{SKmatr}) becomes standard $K$-matrix theory for modular transformations $(a, b, c, d) = (0, -1, 1, d)$ corresponding to the simplest integer and fractional Quantum Hall states with $\sigma^{12}_{xy} = 2/d$.  We also note that there are other ways to arrive at $K$-matrix-like theories and the form Eq.~(\ref{SKmatr}) is not unique, but our solution in the main text is exact and independent of this.

While we do not learn new information about the bulk properties of our model from this $K$-matrix formulation, the connection is still inspiring.  Thus, the form of the $K$-matrix suggests presence of two counter-propagating chiral edge modes, i.e., one non-chiral gapless mode on the boundary.  Our direct analysis and Monte Carlo simulations indeed found power law correlations in observables on the boundary.  We hope that such tractable models can complement the powerful $K$-matrix phenomenology and provide detailed testing grounds for more subtle aspects of edge theories.

%%%%%%%%%%%%%%%%%%%%%%%%%%%%%%%%%%%%%%%%%%%%%%%%%%%%%%%%%%%%%%%%%%%
\subsection{Phenomenological edge theory}
Let us pursue such a $K$-matrix approach and compare with the properties of the quantum Hall edges observed in the main text.  Consider first the \underline{$\sigma^{12}_{xy} = 2$} integer quantum Hall state.  A convenient modular transformation from the physical bosons to gapped quasiparticles is $(a, b, c, d) = (0, -1, 1, 1)$.  Application of Eq.~(\ref{SKmatr}) gives a standard form of the $K$-matrix theory,\cite{Wen_book} which then suggests the following edge theory:
\begin{equation}
S = \int dx d\tau \frac{i}{2\pi} \partial_\tau \varphi_1 \partial_x \varphi_2 + S_{\rm int} ~,
\label{S0}
\end{equation}
where we use Euclidean space-time and orient the edge along the $x$-axis.  Operator $e^{i \varphi_1}$ creates a $\cG_1$ quasiparticle carrying unit charge relative to $\Aext_1$, while $e^{i \varphi_2}$ creates a $\cG_2$ quasiparticle carrying unit charge relative to $\Aext_2$.  This suggests that $e^{i \varphi_1}$ contributes to the physical boson $b_1$, while $e^{i \varphi_2}$ contributes to $b_2$.
[Note that inside the quantum Hall region, the microscopic $\cG_a$ variables are different from the physical $\cJ_a$ variables, and hence $\varphi_a$ are distinct from the microscopic boson phase variables $\phi_a$ used in the Monte Carlo simulations in the main text.  Our crude intuition is that the other side of the boundary can absorb the difference since the vortices are condensed inside the trivial insulator region, and near the boundary we can write schematically $b_a \sim e^{i \varphi_a}$.  Here we do not attempt a microscopic derivation of the edge theory, but rather follow the phenomenological $K$-matrix formalism.\cite{Wen_book, LuVishwanath2012}]
With the above assumptions, we see that at the edge the boson phase fields $\varphi_1$ and $\varphi_2$ behave as conjugate fields,\cite{LuVishwanath2012} similar (up to numerical factors) to fields $\phi$ and $\theta$ in the familiar single-mode Luttinger liquid theory.\cite{Haldane1981, Giamarchi}  This is consistent with our observation that the $b_1$ and $b_2$ power law exponents have opposite trends (see Figs.~\ref{onegood}, \ref{twogood}, and \ref{exponents}), which we discuss further below.

Note that $b_1$ and $b_2$ charge conservation prohibits any cosines of the fields $\varphi_1$ and $\varphi_2$, and the edge is robust.\cite{LuVishwanath2012}
For simplicity, let us consider harmonic interactions of the form\cite{Wen_book}
\begin{equation}
S_{\rm int} = \int\! dx d\tau \frac{1}{4\pi} \left[ U_1 (\partial_x \varphi_1)^2 + U_2 (\partial_x \varphi_2)^2 \right] ~.
\label{Sinteraction}
\end{equation}
We can easily integrate out, say, field $\varphi_2$ and obtain an action for the field $\varphi_1$ only,
\begin{equation}
S_{\varphi_1} = \int\! dx d\tau \frac{1}{4\pi} \left[ U_1 (\partial_x \varphi_1)^2 + \frac{1}{U_2} (\partial_\tau \varphi_1)^2 \right] ~,
\end{equation}
or a similar action for the field $\varphi_2$ only.  We then deduce the scaling dimensions of the boson fields,
\begin{equation}
\Delta[b_1] = \frac{1}{2} \sqrt{\frac{U_2}{U_1}} ~, \quad\quad
\Delta[b_2] = \frac{1}{2} \sqrt{\frac{U_1}{U_2}} ~.
\end{equation}
Hence in this model of the edge, the two scaling dimensions satisfy
\begin{equation}
\Delta[b_1] ~ \Delta[b_2] = \frac{1}{4} ~.
\end{equation}
As discussed in the main text, the numerical results are approximately consistent with the above equation, cf.~Fig.~\ref{exponents}.  There is a slight discrepancy which may be due to the presence of additional terms in Eq.~(\ref{Sinteraction}) or strong finite size effects.

We reiterate that we do not have a microscopic justification of the above edge theory and the specific choices of the interactions.  Nevertheless, the above relation between the scaling dimensions appears to be close to our numerical results for the model studied in the main text.
We suspect that this may be due to a special time-reversal-like symmetry $i \to -i, \phi_1 \to -\phi_1, \phi_2 \to \phi_2$ in the specific model.  While the bulk properties and the edges are robust also without such symmetry, the details of the interactions will differ and there will be no such exact relation.  Nevertheless, we expect similar general trends just from the fact that $\phi_1$ and $\phi_2$ behave as conjugate variables.  This remark applies also to all discussions below.

Consider now the \underline{$\sigma^{12}_{xy} = 2n$} states with $n\geq 2$.  We take $n=2$ as an example.  In this case, any modular transformation that gives gapped quasiparticles is of the form $(a, b, c, d) = (1 + mn, m, n, 1)$ with $m$ an integer; this has $a \neq 0$ and hence our derivation Eq.~(\ref{SKmatr}) gives a somewhat non-standard $K$-matrix theory.  Instead of applying the $K$-matrix formalism, we can obtain a good physical picture of the edge in this case by bringing together two elementary $\sigma^{12}_{xy} = 2$ quantum Hall states and allowing boson hopping between the two ``layers'' labeled $I$ and $II$:
\begin{eqnarray}
S_0 &\!=\!& \int\! dx d\tau \frac{i}{2\pi} \left[\partial_\tau \varphi_1^{(I)} \partial_x \varphi_2^{(I)} + \partial_\tau \varphi_1^{(II)} \partial_x \varphi_2^{(II)} \right] ~~~~~ \\
&\!=\!& \int\! dx d\tau \frac{i}{2\pi} \left[\partial_\tau \varphi_{1+} \partial_x \varphi_{2+} + \partial_\tau \varphi_{1-} \partial_x \varphi_{2-} \right] ~, ~~~~~ \\
\delta S &\!=\!& -\sum_{a = 1, 2} \int\! dx d\tau ~t_a \cos(\varphi_a^{(I)} - \varphi_a^{(II)}) \\
&\!=\!& -\sum_{a = 1, 2} \int\! dx d\tau ~t_a \cos(\sqrt{2} \varphi_{a-}) ~.
\label{Scos}
\end{eqnarray}
Here we have omitted intra- or inter-layer interactions other than the boson tunneling between the layers that preserves the $U(1) \times U(1)$ symmetry.  We have also introduced symmetric and antisymmetric combinations of the phase fields in the two layers,
\begin{equation}
\varphi_{a\pm} = (\varphi_a^{(I)} \pm \varphi_a^{(II)})/\sqrt{2} ~.
\end{equation}
We expect that one of the cosines in Eq.~(\ref{Scos}) is strongly relevant and its coupling will flow to large values and will pin the corresponding phase field.  Note that since $\varphi_{1-}$ and $\varphi_{2-}$ are conjugate variables, only one of them can be pinned.

Let us assume that $\varphi_{1-}$ gets pinned; the conjugate variable $\varphi_{2-}$ then fluctuates wildly.  The remaining fields $\varphi_{1/2,+}$ represent one gapless non-chiral mode, and we can now discuss properties of the resulting edge.
First, we can write the boson fields $b_1$ as
\begin{equation*}
b_1 \sim e^{i \varphi_1^{(I)/(II)}} = e^{i (\varphi_{1+} \pm \varphi_{1-})/\sqrt{2}} = e^{i \varphi_{1+}/\sqrt{2}} \times {\rm const} ~.
\end{equation*}
Thus, we expect power law correlations with the scaling dimension
\begin{equation}
\Delta[b_1] = \frac{1}{4} \sqrt{\frac{U_{2+}^{\rm eff}}{U_{1+}^{\rm eff}}} ~,
\end{equation}
where $U_{a+}^{\rm eff}$ are some effective couplings, again assuming interactions like those in Eq.~(\ref{Sinteraction}) but now for $\varphi_{1+}$ and $\varphi_{2+}$.
On the other hand, the boson fields $b_2$ contain the wildly fluctuating phase $\varphi_{2-}$ in the exponent and hence the $b_2$ correlations are short-ranged.  To obtain power law correlations, we need to consider an object carrying two $b_2$ charges:
\begin{equation}
(b_2)^2 \sim e^{i \varphi_2^{(I)}} e^{i \varphi_2^{(II)}} = e^{i \sqrt{2} \varphi_{2+}} ~,
\end{equation}
which has scaling dimension
\begin{equation}
\Delta[(b_2)^2] = \sqrt{\frac{U_{1+}^{\rm eff}}{U_{2+}^{\rm eff}}} ~.
\end{equation}
This provides a phenomenological explanation of our findings in the specific edge model in the main text, where single-boson fields $b_1$ show power law correlations, while only pair-boson fields $(b_2)^2$ show power law (of course, we can also have situation where the two species are interchanged).  The above scaling dimensions satisfy
\begin{equation}
\Delta[b_1] ~ \Delta[(b_2)^2] = \frac{1}{4} ~.
\end{equation}
We can compare this result to the second panel of Fig.~\ref{exponents}.  The results are approximately consistent, suggesting that our picture of the edge physics is correct and generic.  

We can readily generalize the above argument to the integer quantum Hall case with $n > 2$.  We can also see what is happening in the bulk, again starting with decoupled $\sigma^{12}_{xy} = 2$ ``layers.''  As discussed in the main text, in each layer we have a condensate of bound states of a $b_2$ charge and a vortex in $b_1$.  Borrowing language from layered superconductors (although we assume that each layer can talk to all other layers), vortices in each layer are ``pancake vortices'' and are connected by ``Josephson vortices'' running between each pair of layers.  In the absence of the boson tunneling, the pancake vortices in the layers are uncorrelated and there is no line tension for the Josephson vortices.  When we introduce tunneling, the Josephson vortices acquire line tension and the pancake vortices align near the same location on the 2d plane.  On a coarse-grained scale where the (finite) collection of layers is viewed as a ``fat'' 2d system, such a stack of pancake vortices represents a single vortex in the fat system (indeed, the boson phase winds by the same amount in each layer).  Since we have a $b_2$ charge bound to the pancake vortex in $b_1$ in each layer, we have $n$ such $b_2$ charges bound to this single $b_1$ vortex in the fat system.  This reproduces our picture of the physical origin of the $\sigma^{12}_{xy} = 2n$ integer quantum Hall state as a condensate of bound states of $n$ charges and a vortex.

Turning to the fractional quantum Hall cases, we see that in the \underline{$\sigma^{12}_{xy} = 2/d$} case we can use modular transformation $(0, -1, 1, d)$, and Eq.~(\ref{SKmatr}) gives a standard form of the $K$-matrix theory.  We have an extra factor of $d$ in the edge theory, Eq.~(\ref{S0}), and now $e^{i \varphi_1}$ and $e^{i \varphi_2}$ create quasiparticles carrying charges $1/d$ relative to $\Aext_1$ and $\Aext_2$ respectively, so the microscopic boson fields are represented as $b_a \sim e^{i d \varphi_a}$.  Performing calculations similar to the above, we conclude in this case
\begin{equation}
\Delta[b_1] ~ \Delta[b_2] = \frac{d^2}{4} ~.
\end{equation}

Finally, we can interpret the \underline{$\sigma^{12}_{xy} = 2c/d$} edge by bringing together $c$ more elementary $\sigma^{12}_{xy} = 2/d$ edges.  We again assume that the $b_1$ tunneling between the layers dominates and flows to strong coupling (although this need not always be the case).  We conclude that single-boson $b_1$ correlations are power law, but only ``molecular'' $c$-tupled boson $(b_2)^c$ correlations are power law, and the scaling dimensions are related by
\begin{equation}
\Delta[b_1] ~ \Delta[(b_2)^c] = \frac{d^2}{4} ~.
\end{equation}
We were unable to check these relations numerically because the scaling dimensions $\Delta[b_1]$ were too large to measure.

%%%%%%%%%%%%%%%%%%%%%%%%%%%%%%%%%%%%%%%%%%%%%%%%%%%%%%%%%%%%%%%%%%%
\section{Hamiltonian formulation}
\label{app:H}
Throughout the main text, we worked with the Euclidean action formulation of the model.  The action for the physical currents is local in (2+1)D space-time and is very convenient for analysis.  However, it is natural to ask whether this action can be realized as a path integral of a local Hamiltonian in 2d.\cite{Matthew_Alexei_thanks}  Below we provide an example of such a Hamiltonian.

\begin{figure}
\begin{center}
\includegraphics[width=0.6\linewidth]{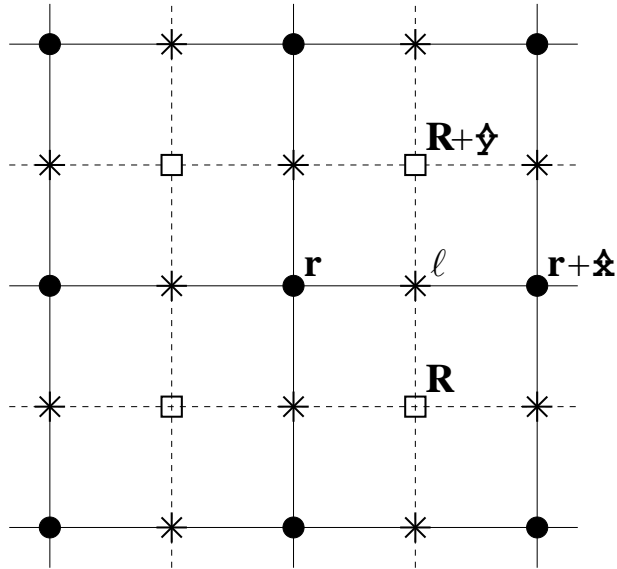}%
\caption{Our Hamiltonian Eq.~(\ref{H}) has U(1) rotors residing on sites $\br$ of the direct square lattice and U(1) rotors residing on sites $\bR$ of the dual lattice.  We also have harmonic oscillators residing on crosses $\ell$ of the links of the two lattices.  The first and second U(1) systems are coupled to the oscillator position and momentum variables respectively as if the latter were gauge fields.\cite{footnoteKitaev}  With appropriate choices of parameters and additional charge-flux couplings, we can induce condensations of bound states of charges and vortices leading to the quantum Hall states discussed in the main text.}
\label{fig:H}
\end{center}
\end{figure}

We first specify the physical Hilbert space.  Our degrees of freedom reside on two inter-penetrating square lattices as shown in Fig.~\ref{fig:H}.  We place quantum U(1) rotors on sites $\br$ of the first square lattice.  The rotors are described by $2\pi$-periodic phase variables $\hat{\phi}_1(\br)$ and conjugate integer number variables $\hat{n}_1(\br)$, with commutation relations $[\hat{\phi}_1(\br), \hat{n}_1(\br')] = i \delta_{\br\br'}$.  We place another set of U(1) rotors, described by $\hat{\phi}_2(\bR)$ and $\hat{n}_2(\bR)$, on sites $\bR$ of the second (dual) square lattice.  Finally, we place harmonic oscillators, described by $\hat{\chi}_\ell$ and $\hat{\pi}_\ell$, on centers of links of the first square lattice, which are also centers of links of the second square lattice, e.g., $\ell = <\br, \br + \hat{\bx}> = <\bR, \bR + \hat{\by}>$ as illustrated in Fig.~\ref{fig:H}.  Here $\hat{\chi}_\ell$ are real-valued coordinate variables and $\hat{\pi}_\ell$ are conjugate momentum variables, $[\hat{\chi}_\ell, \hat{\pi}_{\ell'}] = i\delta_{\ell\ell'}$.  Looking ahead, we will use a path integral containing both $\chi_\ell$ and $\pi_\ell$.  We will view the coordinate variables as fields on the links of the first lattice,
\begin{equation}
\hat{\alpha}_{1j}(\br) \equiv \hat{\chi}_{\br, \br + \hat{\bj}} ~,
\end{equation}
$\hat{\bj} = \hat{\bx}$ or $\hat{\by}$, while we will view the conjugate momentum variables as fields on the links of the second lattice,
\begin{equation}
\hat{\alpha}_{2j}(\bR) = \epsilon_{jk} \hat{\pi}_{\br, \br + \hat{\bk}} ~.
\end{equation}
Here $\epsilon_{xy} = -\epsilon_{yx} = 1$ is the 2d antisymmetric tensor and $<\bR, \bR + \hat{\bj}>$ and $<\br, \br + \hat{\bk}>$ are crossing links.\cite{footnoteKitaev}
Note that in this Appendix we adopt the following notation:  Spatial lattice sites are labeled with bold face, e.g., $\br, \bR$.  Spatial directions are labeled with Roman letters, e.g., $j, k$; space-time directions that appear later will be labeled with Greek letters, e.g., $\mu, \nu$.  Oriented fields residing on spatial links are viewed as spatial vectors and are labeled with bold face, e.g., ${\bm \alpha_1}, {\bm \alpha_2}$.

%%%%%%%%%%%%%%%%%%%%%%xfig output for H figure

Our Hamiltonian is:
\begin{eqnarray}
\label{H}
\hat{H} &=& \hat{H}_{h1} + \hat{H}_{h2} + \hat{H}_{u1} + \hat{H}_{u2} + \hat{H}_\chi + \hat{H}_\pi ~,\\
\hat{H}_{h1} &=& -\sum_{\br, j} h_1 \cos[\nabla_j \hat{\phi}_1(\br) - e_1 \hat{\alpha}_{1j}(\br)] ~,\\
\hat{H}_{h2} &=& -\sum_{\bR, j} h_2 \cos[\nabla_j \hat{\phi}_2(\bR) - e_2 \hat{\alpha}_{2j}(\bR)] ~,\\
\hat{H}_{u1} &=& \frac{1}{2} \sum_\br u_1 \left[\hat{n}_1(\br) + g_1 ({\bm \nabla} \wedge \hat{\bm \alpha}_2)(\br) \right]^2 ~,\\
\hat{H}_{u2} &=& \frac{1}{2} \sum_\bR u_2 \left[\hat{n}_2(\bR) + g_2 ({\bm \nabla} \wedge \hat{\bm \alpha}_1)(\bR) \right]^2 ~,\\
\hat{H}_\chi &=& \sum_\ell \frac{\kappa\, \hat{\chi}_\ell^2}{2} ~, \quad
\hat{H}_\pi = \sum_\ell \frac{\hat{\pi}_\ell^2}{2m} ~.
\end{eqnarray}
Here we introduced various parameters such as boson hopping amplitudes $h_1$ and $h_2$, on-site energies $u_1$ and $u_2$, and oscillator parameters $\kappa$ and $m$.  The hopping terms couple the boson phases and the oscillators as if the latter were ``gauge fields.''  The on-site terms couple the boson numbers and appropriate fluxes of the ``gauge fields'': e.g., flux ${\bm \nabla} \wedge \hat{\bm \alpha}_1 \equiv \nabla_x \hat{\alpha}_{1y} - \nabla_y \hat{\alpha}_{1x}$ is associated with a plaquette of the first lattice, or, equivalently a site $\bR$ of the dual lattice, and is coupled with the boson number $\hat{n}_2(\bR)$ on that site.  The corresponding parameters $e_1, e_2, g_1, g_2$ will be chosen later. 
Here we emphasize that the model is local in the physical variables (i.e., it is not a gauge theory), in the same spirit as Kitaev's toric code model.

We develop imaginary-time path integral by using Trotter decomposition and insertions of unity as follows:
\begin{eqnarray*}
e^{-\delta\tau \hat{H}} &\approx& e^{-\delta\tau (\hat{H}_{u1} + \hat{H}_{h2} + \hat{H}_\pi)} e^{-\delta\tau (\hat{H}_{h1} + \hat{H}_{u2} + \hat{H}_\chi)}\nonumber\\
&&= ~\mathbbm{1}_{\tau + \delta\tau}~ e^{-\delta\tau (\hat{H}_{u1} + \hat{H}_{h2} + \hat{H}_\pi)} ~\mathbbm{1}_{\tau + \frac{\delta\tau}{2}}~ e^{-\delta\tau (\hat{H}_{h1} + \hat{H}_{u2} + \hat{H}_\chi)} ~\mathbbm{1}_{\tau} ~,\\
\mathbbm{1}_{\tau} &=&
\int_{-\pi}^\pi D\phi_1(\br, \tau)
\sum_{n_2(\bR, \tau) = -\infty}^\infty
\int_{-\infty}^\infty D\chi_\ell(\tau) ~\nonumber\\
&&\Big|   \phi_1(\br, \tau), n_2(\bR, \tau), \chi_\ell(\tau) \Big\ra
\Big\la \phi_1(\br, \tau), n_2(\bR, \tau), \chi_\ell(\tau) \Big| ~,\\
\mathbbm{1}_{\tau_\half \equiv \tau + \frac{\delta\tau}{2}} &=&
\sum_{n_1(\br, \tau_\half) = -\infty}^\infty
\int_{-\pi}^\pi D\phi_2(\bR, \tau_\half) 
\int_{-\infty}^\infty D\pi_\ell(\tau_\half) ~\nonumber\\
&&\Big|   n_1(\br, \tau_\half), \phi_2(\bR, \tau_\half), \pi_\ell(\tau_\half) \Big\ra
\Big\la n_1(\br, \tau_\half), \phi_2(\bR, \tau_\half), \pi_\ell(\tau_\half) \Big| ~.
\end{eqnarray*}
Here we used one set of variables on ``integer'' time slices $\tau = {\rm int} \times \delta\tau$ and conjugate variables on ``half-integer'' time slices $\tau_\half \equiv \tau + \delta\tau/2$.  We also arranged the Trotter decomposition so that the pieces of the Hamiltonian act as $c$-numbers on the kets of the above insertions of unity.  Throughout, we omit normalization constants.  The remaining inputs to complete the path integral formulation in the above variables are overlaps such as
\begin{eqnarray*}
\big\la \chi_\ell(\tau + \delta\tau) \big| \pi_\ell(\tau_\half) \big\ra
\big\la \pi_\ell(\tau_\half) \big| \chi_\ell(\tau) \big\ra &=& e^{i \pi_\ell(\tau_\half) [\chi_\ell(\tau + \delta\tau) - \chi_\ell(\tau)]}
= e^{i \alpha_{2k} \epsilon_{kj} \nabla_\tau \alpha_{1j}} ~,\\
%\quad {\rm for~} \ell = <\br, \br+\hat{\bj}>, ~ \\
\big\la \phi_1(\br, \tau + \delta\tau) \big| n_1(\br, \tau_\half) \big\ra
\big\la n_1(\br, \tau_\half) \big| \phi_1(\br, \tau) \big\ra &=&
e^{i n_1(\br, \tau_\half) [\phi_1(\br, \tau + \delta\tau) - \phi_1(\br, \tau)]}
= e^{i \cJ_{1\tau} \nabla_\tau \phi_1} ~,
\end{eqnarray*}
and similarly for the second rotor variables. In the above, $\ell=<\br, \br+\hat{\bj}>$.

In the action, we have phase variables $\phi_1(\br, \tau)$ residing on sites $(\br, \tau)$ of a (2+1)D cubic lattice and $\phi_2(\bR, \tau_\half)$ residing on sites $(\bR, \tau_\half)$ of a dual cubic lattice (in the main text, such space-time points are labeled simply $r$ and $R$).  We also have boson number variables $n_1(\br, \tau_\half)$ and $n_2(\bR, \tau)$, which we can view as residing on temporal links of the first and second (dual) cubic lattices respectively and write as temporal components of boson three-currents, $\cJ_{1\tau}(\br, \tau) \equiv n_1(\br, \tau_\half)$ and $\cJ_{2\tau}(\bR, \tau_\half) \equiv n_2(\bR, \tau + \delta\tau)$.  We introduce spatial current components using an approach familiar in treatments of XY models; namely, we interpret the cosine terms in $\hat{H}_{h1}$ and $\hat{H}_{h2}$ as so-called Villain cosines and write, e.g.,
\begin{eqnarray*}
&& e^{\delta\tau h_1 \cos[\nabla_j \phi_1(\br, \tau) - e_1 \alpha_{1j}(\br, \tau)]} 
 \to \sum_{\cJ_{1j}(\br, \tau) = -\infty}^\infty e^{-\frac{\cJ_{1j}^2}{2 \delta\tau h_1} + i \cJ_{1j} [\nabla_j \phi_1 - e_1 \alpha_{1j}]} ~.
\end{eqnarray*}
We can now integrate over the phase degrees of freedom and obtain current conservation conditions, $\vec{\nabla} \cdot \JJ_1 \equiv \sum_{\mu=x, y, \tau} \nabla_\mu \cJ_{1\mu} = 0$, and similarly for the three-current $\JJ_2$.
Here and below, arrows over symbols denote three-vectors such as $\JJ_1 = (\cJ_{1x}, \cJ_{1y}, \cJ_{1\tau})$, while bold symbols refer to spatial parts such as ${\bm \cJ}_1 = (\cJ_{1x}, \cJ_{1y})$.

We still have the oscillator variables, now labeled ${\bm \alpha}_1(\br, \tau)$ and ${\bm \alpha}_2(\bR, \tau_\half)$ and residing on spatial links of the first and second cubic lattices.  At this point, we could also integrate over these variables and obtain an action in terms of the boson three-currents only.  To facilitate the integration and show the connection with the loop models in the main text, we will first write the on-site terms by introducing auxiliary fields labeled $\alpha_{1\tau}$ and $\alpha_{2\tau}$ residing on the temporal links of the first and second cubic lattices respectively, e.g.:
\begin{eqnarray*}
&& e^{-\frac{\delta\tau u_2}{2} \left[\cJ_{2\tau}(\bR, \tau_\half) + g_2 ({\bm \nabla} \wedge {\bm \alpha}_1)(\bR, \tau_\half) \right]^2} 
 = \int_{-\infty}^\infty d\alpha_{2\tau}(\bR, \tau_\half) e^{-\frac{\alpha_{2\tau}^2}{2 \delta\tau u_2} - i \alpha_{2\tau} \left[J_{2\tau} + g_2 ({\bm \nabla} \wedge {\bm \alpha}_1) \right]} ~.
\end{eqnarray*}
For brevity, we often omit the lattice coordinates on the fields and imply precise geometric relation between objects on different lattices: e.g., an oriented plaquette on one lattice is also associated with a unique oriented bond on the other lattice crossing this plaquette.

Putting everything together, the final action takes the form
\begin{eqnarray*}
&& S[\valpha_1, \valpha_2, \JJ_1, \JJ_2] = 
\sum \left[ \frac{\delta\tau \kappa\, {\bm \alpha}_1^2}{2} + \frac{\alpha_{1\tau}^2}{2\delta\tau u_1} + \frac{\delta\tau {\bm \alpha}_2^2}{2m} + \frac{\alpha_{2\tau}^2}{2\delta\tau u_2} \right]
+ \sum \left[\frac{{\bm \cJ}_1^2}{2\delta\tau h_1} + \frac{{\bm \cJ}_2^2}{2\delta\tau h_2} \right] \\
&&\quad + i \sum \left[ (\nabla_\tau{\bm \alpha}_1) \wedge {\bm \alpha}_2 + g_1 \alpha_{1\tau} ({\bm \nabla} \wedge {\bm \alpha}_2) + g_2 \alpha_{2\tau} ({\bm \nabla} \wedge {\bm \alpha}_1) \right]\nonumber\\
&&\quad + i \sum \left[ e_1 {\bm \cJ}_1 \cdot {\bm \alpha}_1 + \cJ_{1\tau} \alpha_{1\tau} + e_2 {\bm \cJ}_2 \cdot {\bm \alpha}_2 + \cJ_{2\tau} \alpha_{2\tau} \right] ~. 
\end{eqnarray*}
Here the wedge operator is ${\bm v}_1 \wedge {\bm v}_2 \equiv \sum_{j,k} \epsilon_{jk} v_{1j} v_{2k} = v_{1x} v_{2y} - v_{1y} v_{2x}$.
By rescaling ${\bm \alpha}_1 = {\bm \alpha}^\prime_1/e_1$ and similarly for  ${\bm \alpha}_2$, and choosing, e.g., $e_1 = e_2 = 1/g_1 = 1/g_2 = \sqrt{2\pi d/c}$, we obtain essentially the same action as in Eq.~(\ref{preal}) in the main text.  The only difference from the main text is that there are additional local current interactions containing $h_1$ and $h_2$ couplings, and to make the actions identical we only need to take $h_1$ and $h_2$ large.  In particular, the model is in the ``$c/d$'' Quantum Hall phase for sufficiently small $\kappa$ and sufficiently large $m$ and large $u_1, u_2$.
Thus, we have provided a Hamiltonian realization for our Quantum Hall phases.

We can also carry out this derivation when the parameters $h_{1,2}, u_{1,2}, \kappa, m, e_{1,2}, g_{1,2}$ vary in space; in particular, we can study a boundary between Quantum Hall and trivial insulators.
Note that there is significant freedom in how to vary the parameters to achieve different phases even within the specific model.  For example, we can obtain a trivial insulator by taking $e_{1,2}$ and $g_{1,2}$ to be zero while also taking the hopping amplitudes $h_{1,2}$ to be small and on-site potentials $u_{1,2}$ large.
Alternatively, we can take $e_{1,2}$ to be very large while $g_{1,2} \to 0$ and can reach the trivial insulator this way even when the bare hopping amplitudes $h_{1,2}$ are large (in this case, the boson propagation is scrambled by strong phase coupling to the oscillators).  The latter route is closer to the model we used in the main text when discussing a boundary between Quantum Hall and trivial insulators.  Such a boundary model in the present Hamiltonian approach will in general differ from that in the main text.  Indeed, in the Chern-Simons-like piece for the rescaled $\valpha^\prime_{1,2}$ variables,
\begin{equation*}
i \sum [ \frac{(\nabla_\tau {\bm \alpha}^\prime_1) \wedge {\bm \alpha}^\prime_2}{e_1 e_2} + g_1 \alpha_{1\tau} ({\bm \nabla} \wedge \frac{{\bm \alpha}^\prime_2}{e_2}) + g_2 \alpha_{2\tau} ({\bm \nabla} \wedge \frac{{\bm \alpha}^\prime_1}{e_1}) ] ~,
\end{equation*}
the spatially varying $e$ and $g$ couplings can appear non-trivially under spatial derivatives, and the action in general cannot be cast in the form of Eq.~(\ref{JQreal}).  However, we can find a pattern of couplings that will reproduce our boundary model in the main text:  If we take $1/e_a = g_a = \sqrt{c/(2\pi d)}$ for $x\in [x_{aL}, x_{aR}]$ and $1/e_a = g_a = 0$ otherwise, and take the region $[x_{2L}, x_{2R}]$ to be inside the region $[x_{1L}, x_{1R}]$, we eliminate terms with non-desired derivatives of the couplings and can recast the Chern-Simons-like piece for the rescaled $\valpha^\prime_{1,2}$ variables into the form of Eq.~(\ref{JQreal}) with $\eta(R) = c/d$ inside $[x_{2L}, x_{2R}]$ and zero outside.  Thus, we have also provided a Hamiltonian realization of the boundary model used in the main text.

While we universally expect gapless boson correlations on the boundary, detailed aspects can be different for different realizations.  In this paper, we have focused on the crude demonstration of the gaplessness for the specific boundary model in the main text.  In future work, it would be interesting to examine different realizations and systematically explore all aspects of possible edge theories.

\bibliography{bib4twoloops}

\begin{thebibliography}{10}
\bibitem{footnoteKitaev}
{\rm Such toric-code-like structure was suggested to us by Alexei Kitaev}.
\end{thebibliography}
\bibliographystyle{elsarticle-num}
\end{document}